\documentclass[a4paper,fleqn,usenatbib]{mnras}

\usepackage[T1]{fontenc}
\usepackage{ae,aecompl}
\usepackage{float}

\usepackage{graphicx}	% Including figure files
\usepackage{amsmath}	% Advanced maths commands
\usepackage{amssymb}	% Extra maths symbols
\usepackage{caption}
\usepackage{subcaption}
\usepackage{multirow}

\title[DRAO observations of IC 10]{H{\sc i} observations of IC 10 with the DRAO synthesis telescope}

\author[Namumba et al.]{
B.\ Namumba,$^{1}$\thanks{E-mail: brenda@ast.uct.ac.za}
C.\ Carignan, $^{1,2}$ 
T.\ Foster $^{3}$, and 
N.\ Deg $^{1}$
\\
$^1$Department of Astronomy, University of Cape Town, Private Bag X3, Rondebosch 7701, South Africa\\
$^2$Observatoire d$^{\prime}$Astrophysique de l$^{\prime}$Universit$\acute{e}$ de Ouagadougou, BP 7021, Ouagadougou 03, Burkina Faso \\
$^3$Department of Physics and Astronomy, Brandon University, 270-18th Street, Brandon, Manitoba, R7A6A9, Canada
}
\date{Accepted 2019 August 5. Received 2019 August 2; in original form 2018 November 12.}

\pubyear{2019}

\begin{document}
\label{firstpage}
\pagerange{\pageref{firstpage}--\pageref{lastpage}}
\maketitle

\begin{abstract}
H\textsc{i} observations of the nearby blue compact dwarf galaxy IC 10 obtained with the Dominion Radio Astrophysical Observatory synthesis telescope (DRAO), for a total integration of $\sim$1000 hours, are presented. We confirm the NW faint 21 cm H\textsc{i} emission feature discovered in GBT observations. The H\textsc{i} feature has an H\textsc{i} mass of 4.7 $\times 10^{5}$ M$_{\odot}$, which is only $\sim$ 0.6$\%$ of the total H\textsc{i} mass of the galaxy (7.8 $\times 10^{7}$ M$_{\odot}$). In the inner disk, the rotation curve of IC 10 rises steeply, then flattens until the last point where it rises again, with a maximum velocity of 30 km s$^{-1}$. Based on our mass models, the kinematics of the inner disk of IC 10 can be described without the need of a dark matter halo. However, this does not exclude the possible presence of dark matter on a larger scale. It is unlikely that the disturbed features seen in the outer H\textsc{i} disk of IC 10 are caused by an interaction with M 31. Features seen from our simulations are larger and at lower surface density than can be reached by current observations. The higher velocity dispersions seen in regions where several distinct H\textsc{i} features meet with the main core of IC 10 suggests that there is ongoing accretion.

\end{abstract}

%%%%%%%%%%%%%%%%%%%%%%%%%%%%%%%%%%%%%%%%%%%%%%%%%%

%%%%%%%%%%%%%%%%% BODY OF PAPER %%%%%%%%%%%%%%%%%%


%%%%%%%%%%%%%%%%% BODY OF PAPER %%%%%%%%%%%%%%%%%%

\section{Introduction} \label{sa1}

Dwarf galaxies are fundamental laboratories used to constrain the impact that structural and intrinsic properties have on the evolution of galaxies. Investigating the physical processes occurring in dwarf galaxies is therefore crucial for understanding a wide range of astrophysical phenomena, from star formation mechanisms to the way galaxies are connected to their environment. Among the different dwarf galaxies, blue compact dwarf galaxies (BCDs) are the most fascinating and puzzling objects. While star forming dwarf galaxies are generally known to be inefficient at transforming their gas into stars \citep{1538-3881-136-6-2782}, a different case is observed for BCDs. BCDs are known to be undergoing violent bursts of star formation  \citep{1970ApJ...162L.155S,1994AJ....107.2021M,2003ApJS..147...29G,2004ApJ...616..752L,2006ApJ...639..157W}. However it remains unclear what mechanisms are responsible for the ignition of the star bursts in these systems.

Whilst studies have shown that gravitational interaction, mergers, or accretion can explain the burst of star formation in certain BCDs \citep{1998AJ....116.2363W,2001A&A...371..806N,2008MNRAS.388L..10B,2013ApJ...779L..15N,2014MNRAS.442.2909C}, the absence of nearby companions around most BCDs suggests that mergers or interactions are unlikely starburst triggers. This, however, does not completely rule out possible interactions with companions or H\textsc{i} clouds that may be too faint to be detected and may have been missed in earlier studies of BCDs. 

The neutral hydrogen atomic (H\textsc{i}) gas has been known to be an ideal tracer for the starburst triggers in galaxies \cite{1993AJ....105..128T,1995ApJS...99..427T,1996ApJS..102..189T,1996ApJS..107..143T}. This is because external disturbances are detected easily in H\textsc{i} 21 cm-line images in comparison to the optical images. From the H\textsc{i} morphology, signatures of interactions and mergers can easily be seen in the form of bridges or tidal tails \citep{2013ApJ...779L..15N}. Bridges are often associated with early stages of mergers or on-going interactions while tidal tails are often linked to past close encounter events with other galaxies. In addition to this, progressive stages of interactions can be traced from H\textsc{i} kinematical disturbances. 

IC 10 (UGC 192) located at a distance of 0.7 Mpc \citep{2012AJ....144..134H} is the only known BCD galaxy in the Local Group. Its proximity makes it an excellent candidate for exploring starburst triggers in BCD galaxies. IC 10 is located at low galactic latitude (l=118$^{\circ}$.95,b=-3$^{\circ}$.33) where  the  foreground  reddening  is  expected  to  be  large.  In addition,  the  internal  reddening  of  IC  10  may  vary  spatially due to its strong star-forming activity. Therefore, it is difficult to determine reliably the reddening and distance of IC 10 \citep{2009ApJ...703..816K}. The presence of so many WR stars \citep{1998ARA&A..36..435M} and high H${\alpha}$ luminosity suggest that the starburst in IC 10 must have occurred $\sim$ 10 Myrs ago \citep{2001A&A...370...34R}. The basic parameters of IC 10 are given in Table \ref{tab:template}. The first H\textsc{i} observations of IC 10 with the single dish 100 m Effelsberg telescope revealed H\textsc{i} extending 80$^\prime$ in diameter \citep{1979A&A....75..170H}. Early interferometric H\textsc{i} observations showed that the outer regions of IC 10 were rotating counter-clockwise with respect to a regularly inner rotating disk \citep{1989A&A...214...33S,1998AJ....116.2363W}. \citet{2008AIPC.1035..156M} obtained a deep H\textsc{i} mosaic of IC 10 using the Westerbrook interferometer and found a complex and disturbed H\textsc{i} distribution covering almost 1 deg$^{2}$. H\textsc{i} observations with the Green Bank Telescope revealed a new extended faint feature stretching 1.3 degrees north-west of IC 10 \citep{2013ApJ...779L..15N}. Detailed analysis of these observations combined with high spatial resolution VLA data (LITTLE THINGS) showed that the H\textsc{i} disk of IC 10 is much more extended than previously reported \citep{2014AJ....148..130A} if the faint features seen with the GBT  are real (see Section \ref{sa5}). 

Different mechanisms have been proposed to explain starburst triggers and the observed disturbed morphology and kinematics in IC 10. \citet{1989A&A...214...33S} suggested that the observed complex morphology of IC 10 could be the result of IC 10 colliding with the intergalactic medium. They, however, highlighted that given the similarity of IC 10 with other dwarf galaxies, it is likely that the H\textsc{i} observed in this galaxy is primordial gas which is still in a collapse phase. \citet{1998AJ....116.2363W} analyzed high-resolution H\textsc{i} data of IC 10 and concluded that the observed extended and counter-rotating distribution of gas in IC 10 was due to accretion. They explained that the features seen in IC 10 correspond to what would be seen for a galaxy experiencing later stages of formation via the ongoing infall of primordial material. Recent H\textsc{i} studies of IC 10 suggest that an interaction or merger with another unknown low surface density dwarf galaxy is the most probable explanation for the observed morphology of this galaxy \citep{2013ApJ...779L..15N,2014AJ....148..130A}. However, \citet{2015MNRAS.454.1000G} argued that the lack of streams or shells, or any disturbances in the stellar component of IC 10 strongly indicates that it is unlikely that IC 10 has undergone an interaction with an unknown companion. \citet{2013ApJ...779L..15N} used a Gravitylab N-body integrator code to perform orbit calculations to determine if the origin of the new H\textsc{i} feature near IC 10 was due to an IC 10 -- M  31 interaction. Their results show that the H\textsc{i} extension is inconsistent with the trailing portion of the orbit, making it unlikely that an M 31 -- IC 10 tidal interaction can explain the origin of this feature. We will explore this further in Section \ref{s9}. 

The majority of published H\textsc{i} synthesis observations on IC 10 discuss the disturbed morphology and kinematics of this galaxy based on observations only. The orbital simulations done by \citet{2013ApJ...779L..15N} focused on the origin of the new H\textsc{i} feature and not on the entire disturbed morphology of IC 10. To advance our understanding on possible mechanisms responsible for the observed H\textsc{i} structure of IC 10, we have combined these new DRAO H\textsc{i} observations  with simulations. The DRAO observations allow us to study the extended H\textsc{i} morphology and kinematics while the simulations provide detailed investigation of possible interactions between IC 10 and M 31. The outline of this paper is as follows. In Section \ref{sa2} we discuss the observations and data reduction. In Section \ref{sa3} we present the results of the H\textsc{i} distribution of IC 10. In Section \ref{s44} we discuss the tilted ring and mass model results of the main disk of IC 10. Then in Section \ref{s9} we provide simulations to see if M 31 could be responsible of the outer H\textsc{i} structure of IC 10. The summary and conclusions of this work are given in Section \ref{s66}.
\begin{table}
\scriptsize

\caption{\small Basic parameters of IC 10}
\begin{minipage}{\textwidth}
\begin{tabular}{l@{\hspace{0.90cm}}c@{\hspace{0.5cm}}}   
\hline

Parameter & \\
                         %&(sec)       &(Mpc) & (mag) & ($L_{\odot}$) &($\rm M_{\odot}yr^{-1}$)  \\
                %~~~~~~(1)    &   (2)        &  (3)   \\         
   
\hline \hline  
%\multicolumn{6}{@{} p{8.5 cm} @{}}{\footnotesize{\hspace{3cm} VLT/NACO DATA}}\\\\
Morphology  & BCD$^{b}$\\
Right ascension (J2000) &00:20:24.6$^{a}$\\
Declination (J2000)&+59:17:30.0$^{a}$\\
Distance (Mpc) & 0.7$^{e}$\\
M$_{V}$ (Mag)& -16.3$^{e}$\\
$\alpha^{-1}$ (kpc) & 0.40 $\pm$ 0.01$^{d}$\\
V$_{\text{heliocentric}}$ (km s$^{-1}$) & -348.0$^{c}$\\
Optical PA$^{(\circ)}$ &-38.0$^{e}$\\
Optical inclination$^{(\circ)}$ & 41.0$^{e}$\\
Total HI mass (M$_{\odot}$)& 8.0 $\times10^{7}$$^{c}$\\
A$_{V}$ (mag)&4.3$^{f}$\\
Galactic latitude($^{\circ}$)&118.95$^{g}$\\
Galactic longitude($^{\circ}$)& -3.33$^{g}$\\
proper motion (RA (kms$^{2}$))& -122 $\pm$ 31\\
proper motion (DEC (kms$^{2}$))& -97 $\pm$ 27\\
\hline     \\
\multicolumn{2}{@{} p{9 cm} @{}}{\footnotesize{\textbf{Notes.} Ref\,(a) \cite{1991rc3..book.....D}; (b) \cite{2001A&A...370...34R}; (c) \cite{2014AJ....148..130A}; (d) \cite{0067-0049-162-1-49}; (e) \cite{2012AJ....144..134H}; (f) \cite{1998ApJ...500..525S}; (g) \cite{2009ApJ...703..816K}}}
\label{coords_table}
 
\end{tabular}   
\label{tab:template}
\end{minipage}
\end{table}

\begin{table}
\scriptsize

\caption{\small DRAO Synthesis Telescope observational set-up}
\begin{minipage}{\textwidth}

\begin{tabular}{l@{\hspace{0.8cm}}c@{\hspace{0.5cm}}c@{\hspace{0.4cm}}c@{\hspace{0.4cm}} }   
\hline

Obs.&Field center &  Beam parameter & 1-$\sigma$ Noise at \\
data&coordinates&&field center\\
&J2000&$\theta_{maj}(^{\prime \prime}) \times \theta_{min}(^{\prime \prime})$& (mJy/beam)\\
                         %&(sec)       &(Mpc) & (mag) & ($L_{\odot}$) &($\rm M_{\odot}yr^{-1}$)  \\
                %~~~~~~(1)    &   (2)        &  (3)   \\         
   
\hline \hline  
%\multicolumn{6}{@{} p{8.5 cm} @{}}{\footnotesize{\hspace{3cm} VLT/NACO DATA}}\\\\
20 Feb., 1997 &0:20:23.1, +59:19:08.4&67.48 $\times$ 58.98 & 13.2\\
9 Sept., 2017 &0:20:17.3, +59:14:14&67.63 $\times$ 59.13 & 11.5 \\
9 Sept., 2017 &0:20:17.3, +59:22:14.0& 68.18 $\times$ 59.98 & 11.1\\
1 Oct., 2018 &0:22:28.0, +60:25:09.0& 66.19 $\times$ 59.11 & 11.1\\
5 Apr., 2019 &0:16:12.8, +60:05:29.8& 66.40 $\times$ 58.97 & 11.2\\
5 Apr., 2019 &0:17:32.8, +59:49:35.6& 68.69 $\times$ 58.95 & 10.9\\
5 Apr., 2019 &0:14:49.1, +60:21:18.3& 69.31 $\times$ 59.39 & 11.0\\
\hline     
 
\end{tabular}   
\label{tab:template1}
\end{minipage}
\end{table}

\begin{table}
\scriptsize
\captionsetup{width=8.5cm}
\caption{\small Observational parameters for the full resolution DRAO Synthesis Telescope mosaic}
\begin{minipage}{\textwidth}
\begin{tabular}{l@{\hspace{0.70cm}}c@{\hspace{0.05cm}}c@{\hspace{0.05cm}}}   
\hline

Parameter & Value&\\
                         %&(sec)       &(Mpc) & (mag) & ($L_{\odot}$) &($\rm M_{\odot}yr^{-1}$)  \\
                %~~~~~~(1)    &   (2)        &  (3)   \\         
   
\hline \hline  
%\multicolumn{6}{@{} p{8.5 cm} @{}}{\footnotesize{\hspace{3cm} VLT/NACO DATA}}\\\\
Total length of observation &864 hrs (6 fields)&\\
Integration length of archived data &144 hrs (1 fields)&\\
Frequency center of band&1422.0643 MHz &\\
Number of velocity channels& 256&\\
synthesized beam & 59$^{\prime \prime} \times$ 68$^{\prime \prime}$\\
Velocity resolution& 2.64 km s$^{-1}$&\\
Number of spatial pixels&1024&\\
Pixel angular size & 18$^{\prime \prime}$\\
1$\sigma$ RMS noise (single channel)& 0.8 Kelvin (5.3 mJy/beam)\\
Measured 3$\sigma$ column density over 16 km s$^{-1}$&3.1 $\times$ 10$^{19}$cm$^{-2}$\\
\hline     \\
\multicolumn{3}{@{} p{8 cm} @{}}{\footnotesize{\textbf{}}}
\label{coords_table}
 
\end{tabular}  

\end{minipage}
\label{tab:template2}
\end{table}  
%\vspace{-5cm}
\section{DRAO observations and data reduction}\label{sa2}

%\textcolor{red}{Give more information on the EBHIS data in section 2. What is the noise level, TYLER}
%\textcolor{red}{Give more information on how the DRAO and EBHIS data combined. This is a nontrivial %step.TYLER}

The primary observations for this study were made in the 21~cm line of neutral hydrogen with the Synthesis Telescope at the 
DRAO. This telescope is an East-West interferometer consisting of seven 9~m diameter dishes spaced 
variously across a maximum baseline of 617.2~m. This longest baseline achieves a synthesized half-power beamwidth of 49$^{\prime\prime}$(EW) $\times $49$^{\prime\prime}$/sin~$\delta$(NS) with uniform weighting, although in the H\textsc{i} line we chose natural weighting (a Gaussian taper in the $u,v$ plane) to increase the sensitivity at the slight expense of resolution (58$^{\prime\prime} \times$ 58$^{\prime\prime}/\textrm{sin}\delta$). Further specifications and capabilities of the DRAO can be found in \citet{2000A&AS..145..509L} and \citet{2010ASPC..438..415K}. Other observing frequencies of IC~10 were in a 30~MHz continuum band centered at 1420~MHz ($\lambda$ = 21~cm), and a 2~MHz band at 408~MHz ($\lambda$ = 74~cm). The telescope also produces fully calibrated and instrumentally-corrected Stokes U,Q,V polarization maps, observed in four 7.5~MHz-wide sub-bands centred at 1420~MHz; however the radio polarization and continuum data for IC~10 are not presented here.

Six full synthesis fields were observed on and immediately around IC~10, each 12 hours (per set of spacings) times 12 days. The spectrometer was set to a 2~MHz-wide band in the line with 256 channels spaced by $\Delta v=$1.65~km~s$^{-1}$ per channel. The band was centered on a heliocentric velocity of $v_{\textsc{hel}}=-$350~km~s$^{-1}$. The synthesized beam of these observations is 59$^{\prime\prime} \times$ 68$^{\prime\prime}$, and the velocity resolution is 2.64~km~s$^{-1}$. The bandwidth and these resolution parameters produce a 1$\sigma$-RMS noise in each (empty) channel of 1.5-1.7~K. To increase sensitivity to the extended H\textsc{i} disk of IC~10, another full-synthesis field with the same pointing centre and observing parameters was obtained from the DRAO archive (observed in 1997). Table \ref{tab:template1} lists the field centers, synthesized beamwidths and typical RMS noise for the seven total individual fields (noise figures quoted are measured after continuum subtraction).

Prior to mosaicking the seven fields together, we perform standard data processing steps on each developed for 
the Canadian Galactic Plane Survey \citep[described in][]{2003AJ....125.3145T}. Channels that are free of line-emission 
are averaged at both the low and high velocity ends of the band to produce continuum-only maps, and a 
map that is linearly-interpolated at frequencies between these two is subtracted from each channel in the data cube. To 
calibrate the flux scale of each continuum-subtracted cube, we compare point sources in the mean of both 
continuum end-channel maps to those in the 30~MHz continuum map of each field (these maps are first CLEANed 
around the stronger sources). The observed 30~MHz wide continuum maps are flux calibrated against several 
strong sources observed by the ST \citep[e.g. 3C~48, 3C~286; see Table 1 of][for sources and fluxes]{2003AJ....125.3145T}. 
An error-weighted mean of the flux ratio $S_{1420 cont}/S_{H\textsc{i} cont}$ for all point-sources above a cutoff 
level (20~mJy~beam$^{-1}$) is obtained; this sample is further trimmed by rejecting sources $\pm$2$\sigma$ 
away from the mean. Typically, 20-30 sources remained in the sample for each field. The uncertainty introduced to the flux by this step is $\sim$5\%. 

Each synthesis field is trimmed to the half-power width of the primary-beam of the individual antennae 
(107$^{\prime} \times 107^{\prime})$, and corrected from the primary beam pattern prior to mosaiking. The seven fields were mosaicked together with each weighted proportionally to the inverse of their 
relative RMS noise, and points across the field are assigned a weight proportional to the primary-beam pattern squared. The final step was to recover missing spatial structures (aka {}``short-spacings'') that the DRAO ST is not sensitive to (larger than about 45$^{\prime}$). We use the Effelsberg H{\textsc i} Survey \cite{2016A&A...585A..41W}, a single-dish survey with a beam of 10.82 arcminutes and a noise per 1.29~km~s$^{-1}$-wide channel of $\Delta T_B=$0.09~K. Using this survey gives excellent spatial-frequency overlap with the DRAO ST, providing a smooth transition between structures imaged by the interferometer and the 100-m Effelsberg dish. We add these data together in the map-plane, first by smoothing ST maps to the Effelsberg beam, and then adding the difference between the corresponding EBHIS map and this smoothed ST map back to the high-resolution ST-only data. The resulting short-spacings-added map contains no negative {}``bowls'' around any extended emission strong or weak down to the level of the noise, giving confidence not only in our simple map-plane approach to adding in structures but in our flux calibration as well.
Once stacked the measured 1$\sigma$ noise at the centre of the final 7-field short-spacings-added mosaic is 5.3~mJy/beam, or $\pm$0.8~K per (empty) channel for the full resolution of 59$^{\prime\prime} \times 68^{\prime\prime}$. Table \ref{tab:template2} shows the parameters for the final mosaic cube of IC 10. 

\section{H{\sc i} distribution and kinematics of IC 10}\label{sa3}
The H{\sc i} maps of IC 10 were created using the H\textsc{i} source finding algorithm, SoFiA \citep{2015MNRAS.448.1922S}. The ability of SoFiA to search for emission on multiple scales while taking into account artifacts and variations in noise allow us to deal with the complex H\textsc{i} structure of IC 10 with much ease. The smooth + clip source detection algorithm was used to search for emission by setting a detection limit of 2$\sigma$ which corresponds to a column density limit of 10$^{19}$ cm$^{-2}$. A detailed description of this software and how it is used is given in \citet{2015MNRAS.448.1922S}. 
\subsection{The H{\sc i} global profile }\label{s4}
\begin{figure}
  % Maximum length
  \includegraphics[width=3.5in]{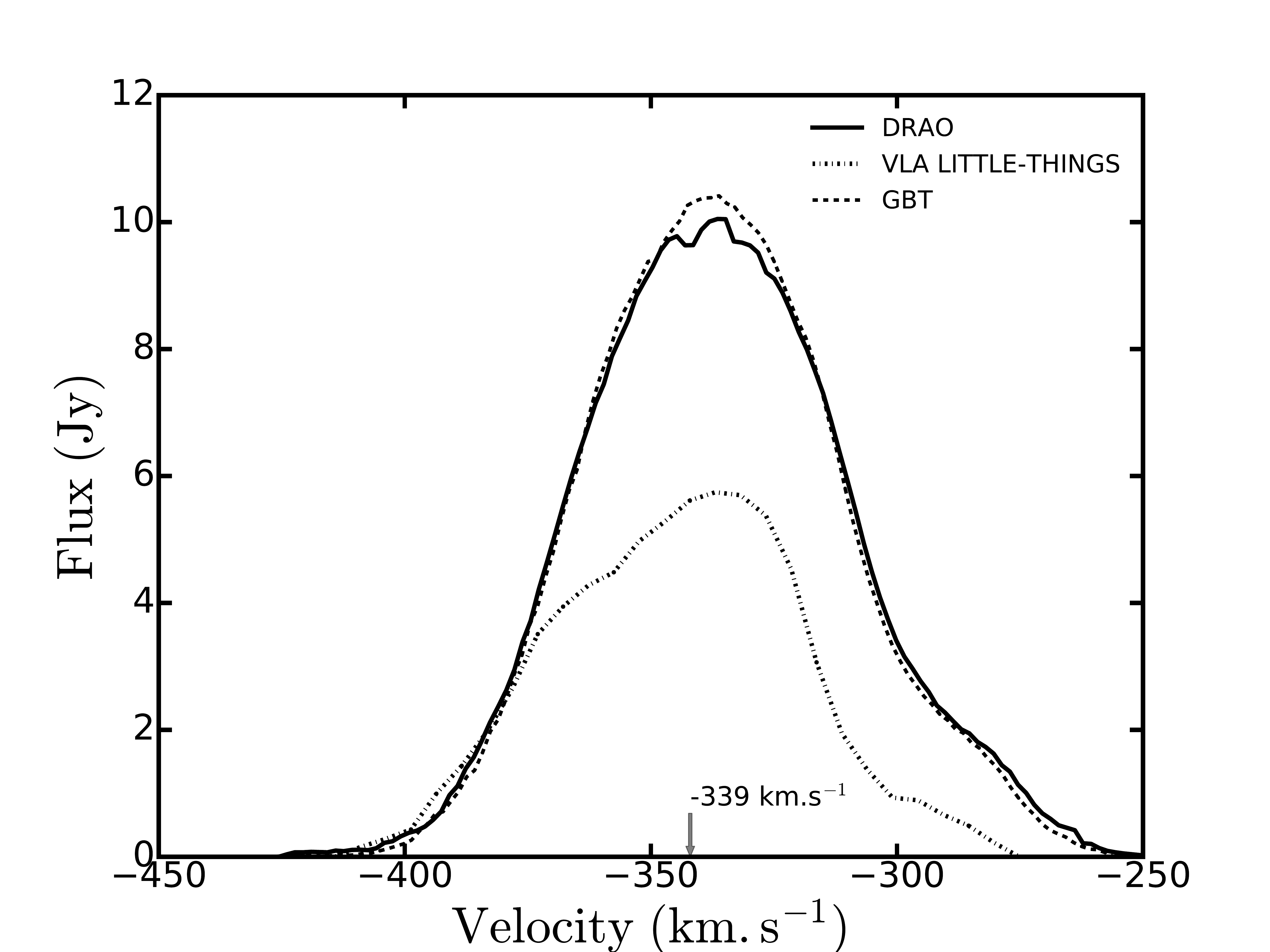}\hspace{-1.2em}%
   \caption{Comparison of the H\textsc{i} global profile of IC 10 from the DRAO map (solid line, this work), from the VLA LITTLE THINGS (dash-dotted line, \citet{2012AJ....144..134H}), and from the GBT map (dotted line, \citet{2013ApJ...779L..15N}.}  
 \label{Fig:a}
\end{figure}

The H\textsc{i} global profile, derived from a primary beam corrected data cube is presented in Figure \ref{Fig:a}. We compare the DRAO global profile to the GBT \citep{2013ApJ...779L..15N} and VLA LITTLE THINGS \citep{2012AJ....144..134H} global profiles. The resulting profile is asymmetric with more H\textsc{i} around the southern extension ($\sim$-300 km s$^{-1}$ to -250 km s$^{-1}$) where the HI is on scales $>$ 15$^{\prime}$ not seen by the VLA. From the profile, a mid-point velocity of -339 $\pm$ 6 km s$^{-1}$ at 50$\%$ level is found, which is a better representation of the systemic velocity than the intensity weighted mean because of the asymmetry of the profile. This value is comparable to the value of  -338 km s$^{-1}$ found by \citet{2013ApJ...779L..15N}. The profile widths calculated at the 20 and 50 per cent levels of the peak flux density are $\Delta$V$_{20}$ = 96 $\pm$ 3 km s$^{-1}$ and $\Delta$V$_{50}$ = 61 $\pm$ 3 km s$^{-1}$. A total flux of 675 $\pm$ 6 Jy km s$^{-1}$ is estimated from the global profile. The total H\textsc{i} mass is calculated as
\begin{equation}
M_{H\textsc{i}} = 2.36 \times 10^{5} \bigg(\frac{D}{\text{Mpc}}\bigg)^2 \int \text{FdV}
\end{equation}
where $\int \text{FdV}$ is the source total flux in units of Jy km s$^{-1}$ and $D$ is the distance in Mpc. Adopting a distance of 0.7 Mpc \citep{2012AJ....144..134H}, a total
H\textsc{i} mass of (7.8 $\pm$ 0.07)$\times$ 10$^{7}$ M$_{\odot}$ is calculated. This mass is larger by $\sim$ 36$\%$ than the value of (5 $\pm$ 0.046) $\times 10^{7}$ M$_{\odot}$ measured from the VLA data \citep{2014AJ....148..130A} and is smaller by $\sim$ 1$\%$ than the value of (7.9 $\pm$ 0.08) $\times 10^{7}$ M$_{\odot}$ measured from the GBT data \citep{2013ApJ...779L..15N}. All masses are calculated using our adopted distance. This result shows that no flux has been missed by the DRAO observations. This is because the compact configuration of DRAO is sensitive to extended H\textsc{i} emission that is missed by the VLA and detected with the single dish GBT. 
\subsection{H{\sc i} distribution}
The H\textsc{i} intensity map extracted from the natural weighted H\textsc{i} data cube is shown in Figure \ref{Fig:b}(a). As has been seen in previous H\textsc{i} interferometric observations \citep{1989A&A...214...33S,1998AJ....116.2363W,2014AJ....148..130A}, the H\textsc{i} distribution of IC 10 is characterized by several distinct H\textsc{i} features. These features include a southern plume, a rotating arm, and spurs north-west and north-east of IC 10. It is important to note that although all these features were previously reported, the DRAO observations are sensitive to large scale structures
and detect these features out to a much larger extent. For instance, the last VLA H\textsc{i} observations of IC 10 measure the southern plume out to 17$^{\prime}$ \citep{2012AJ....144..134H} while our DRAO observations measure this feature out to 32$^{\prime}$. Figure \ref{Fig:c} shows the total column density map of the central parts of IC 10 superposed on a WISE 3.4 $\mu$m band map, which is sensitive to light from the old stellar populations. The brightest H\textsc{i} peak coincides with the stars.

\subsection{H{\sc i} kinematics}
\begin{figure*}
  % Maximum length
  \subcaptionbox{Intensity map\label{fig1:a}}{\includegraphics[width=2.4in]{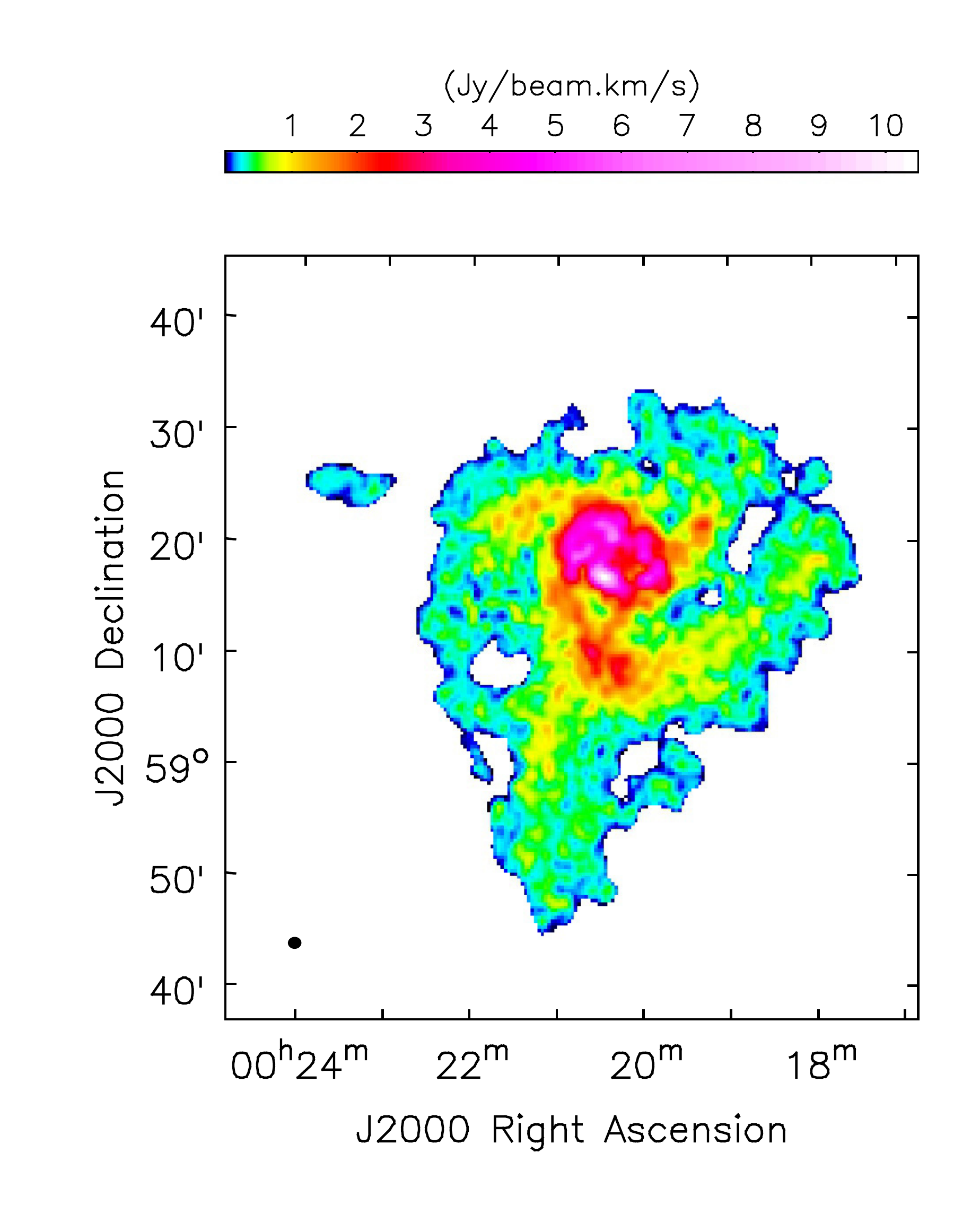}}\hspace{-1.2em}%
  \subcaptionbox{Velocity map\label{fig1:b}}{\includegraphics[width=2.4in]{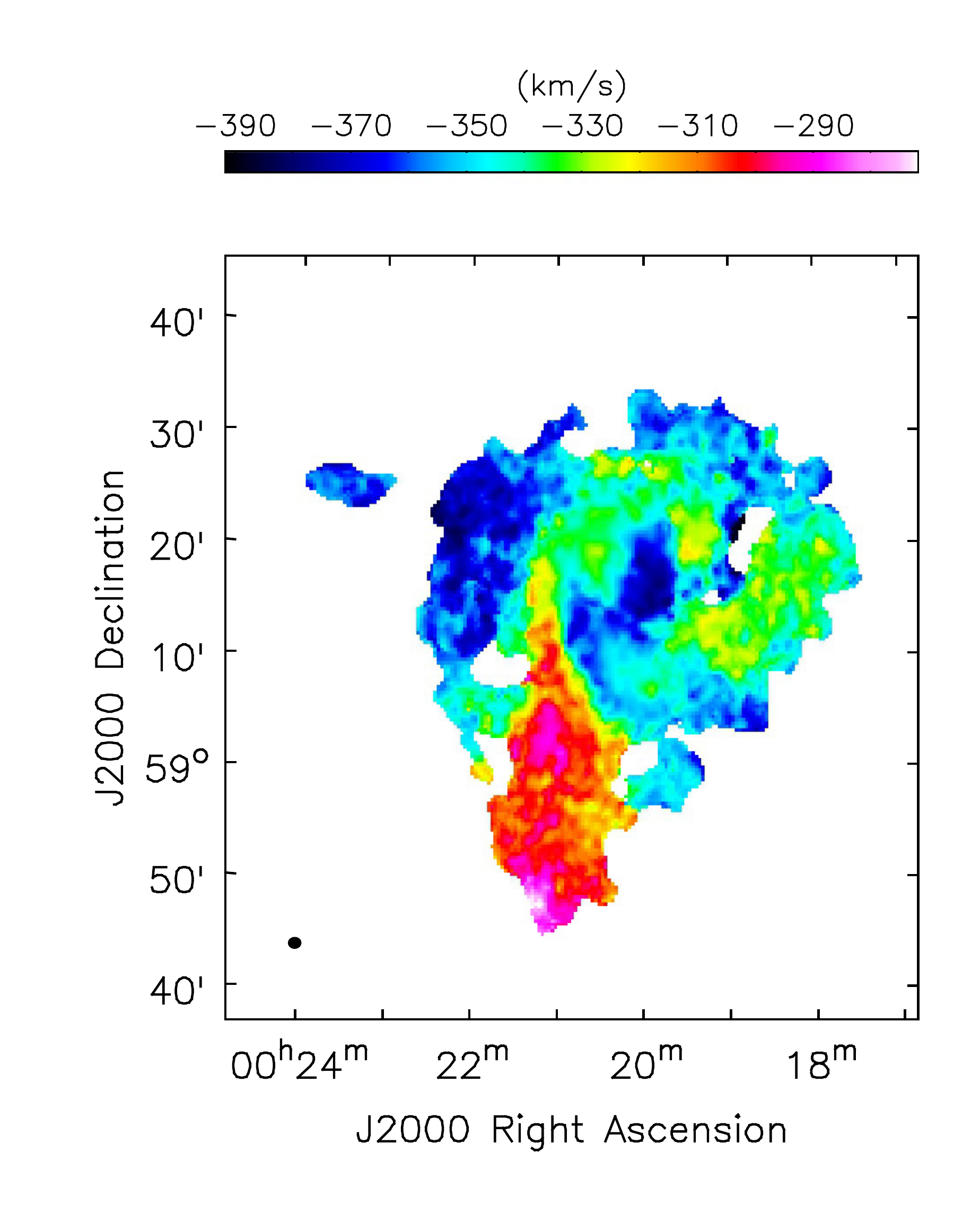}}\hspace{-1.2em}
  \subcaptionbox{Dispersion map\label{fig1:c}}{\includegraphics[width=2.4in]{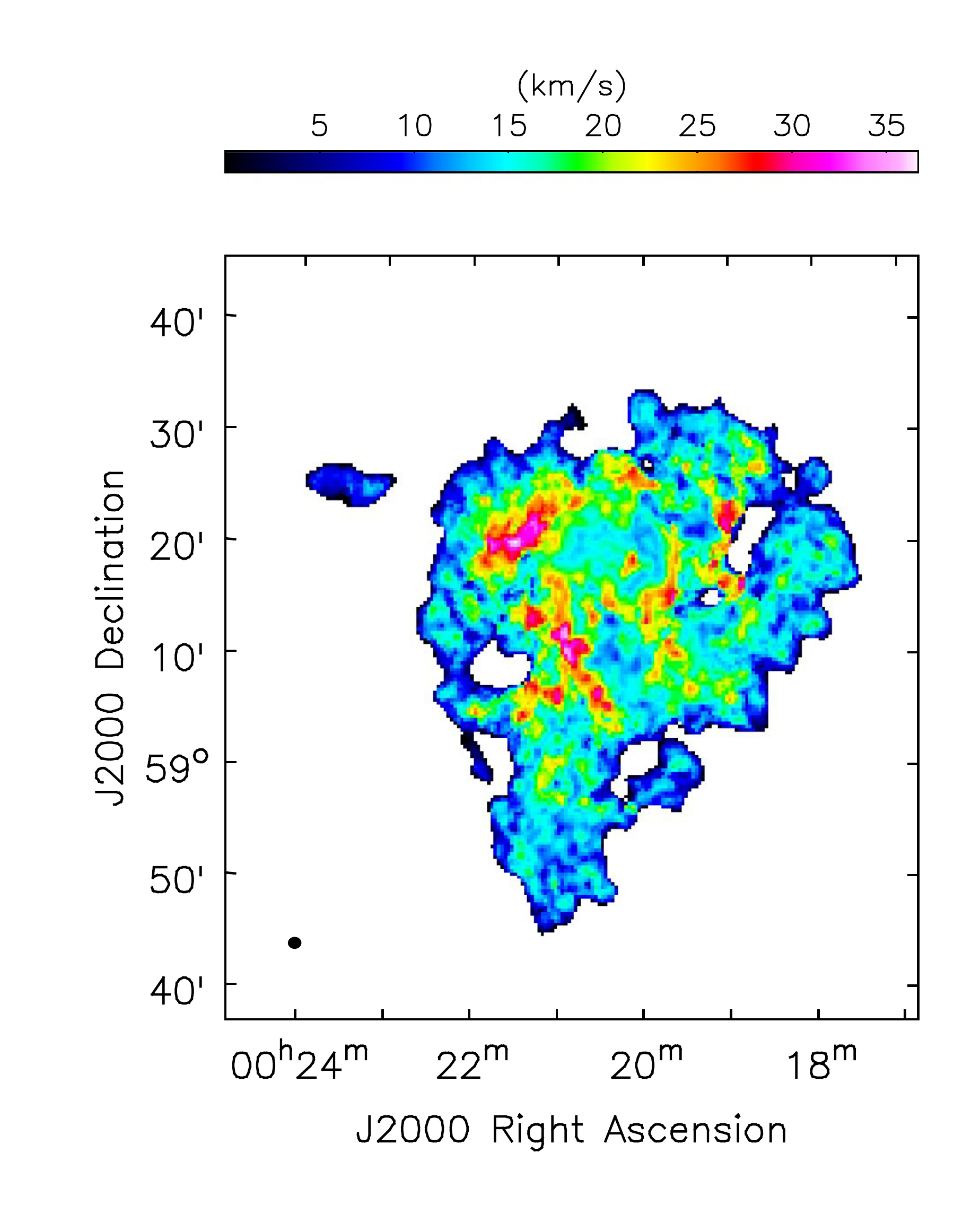}}\\
  \subcaptionbox{PV diagram \label{fig1:d}}{\includegraphics[width=2.8in]{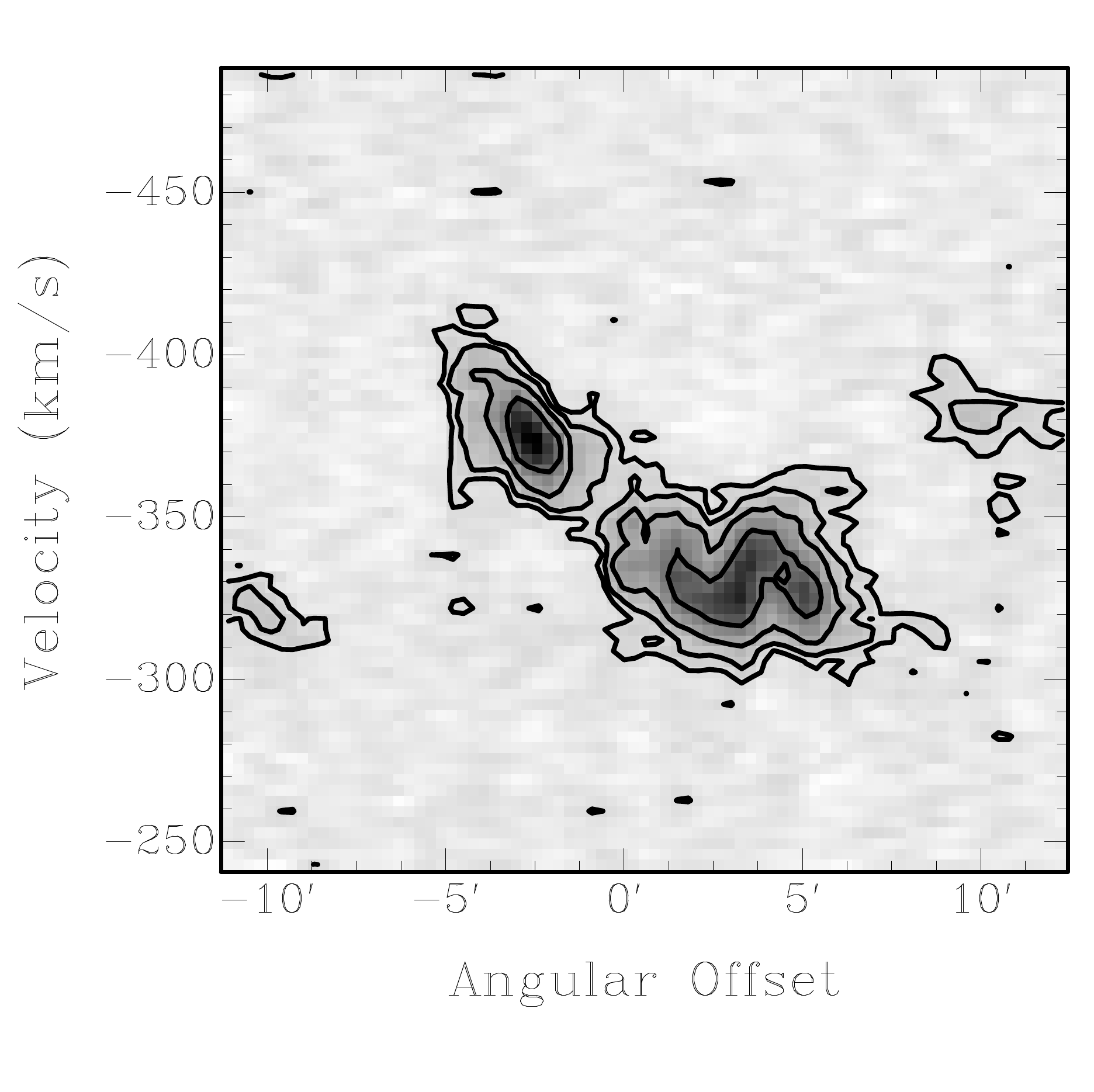}}

  \caption{Moment maps of IC 10 created using the SoFiA software. Figure \ref{Fig:b}(a) shows the intensity map, Figure \ref{Fig:b}(b), the velocity field map, Figure \ref{Fig:b}(c) is the dispersion map, and Figure \ref{Fig:b}(d) is the position-velocity diagram of IC 10 along the major axis. The contours are at 1, 2, 4 , 8, 16 $\times$ 3$\sigma$. where 1$\sigma$ = 0.0053 Jy/beam.}  
  \label{Fig:b}
\end{figure*}

\begin{figure}
  % Maximum length
  \includegraphics[width=3.4in]{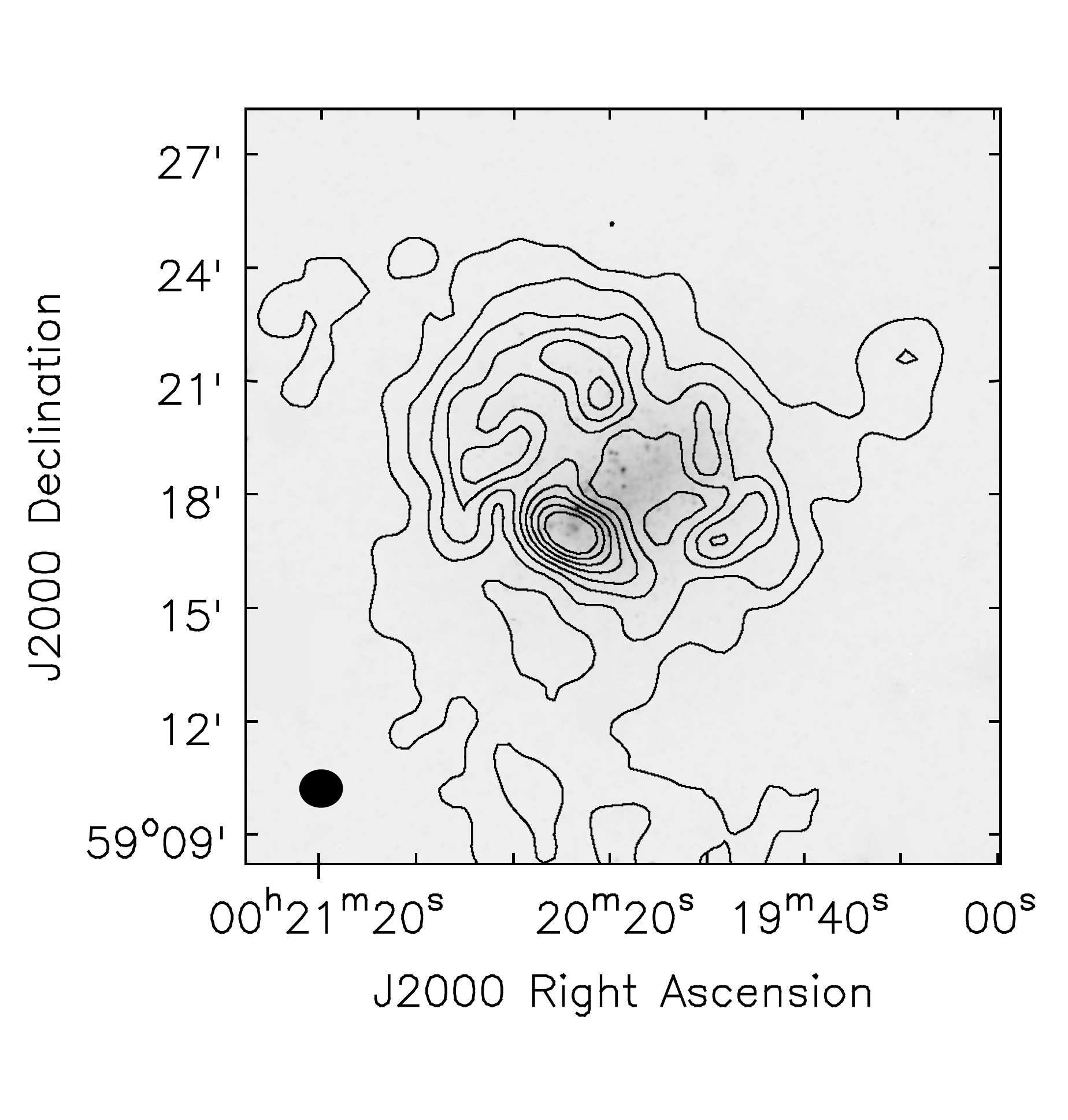}\hspace{-1.2em}%
   \caption{Integrated H\textsc{i} column density contours of IC 10 from the DRAO data overlaid on a WISE 3.4 $\mu$m map. The H\textsc{i} column density contours are at 1, 2, 3, 4, 5, 6, 7, and 8 $\times$ 10$^{20}$ cm$^{-2}$. The synthesized beam is shown in the bottom left corner.} 
 \label{Fig:c}
\end{figure}

The velocity field and dispersion maps of IC 10 are shown in Figure \ref{Fig:b}(b) and \ref{Fig:b}(c) respectively. As noted in previous H\textsc{i} observations, IC 10 is characterized by a distorted velocity field. The southern plume is seen to have velocities different from the rest of the galaxy. A closer look at the central region of the velocity field map shows that the central disk of IC 10 behaves like a rotating disk (see below, Figure \ref{Fig:f}(a)). On the other hand, the spurs and the southern plume are counter-rotating with respect to the central disk. The velocity dispersion of IC 10 varies from $\sim$ 10 km s$^{-1}$ to $\sim$ 35 km s$^{-1}$. High velocity dispersion is seen in regions where different velocity structures meet with each other (Figure \ref{Fig:b}(b)). This indicates that the southern plume, the west and east extensions are probably accreting on the main body of IC 10 as suggested by \citet{1998AJ....116.2363W}, and this could explain the recent starburst that occurred $\sim$ 10 Myr ago \citep{2001A&A...370...34R}. 

\subsection{GBT new H{\sc i} feature of IC 10} \label{sa5}

One of the aims of this study was to verify the existence of the NW faint 21 cm H\textsc{i} feature discovered in the GBT observations and detected at a column density N$_{\text{H\textsc{i}}}$ $\sim$ 7 $\times$ 10$^{17}$ cm$^{-2}$ \citep{2013ApJ...779L..15N}. Verifying whether this feature is real or not is crucial for our interpretation of the H\textsc{i} data of IC 10. If this feature is an H\textsc{i} cloud it could explain the very disturbed and highly extended H\textsc{i} disk of IC 10 (through possible interaction, including tidal stripping and merging). If it is a new satellite galaxy of IC 10, the intermediate spatial resolution of DRAO will allow us to explore its kinematics and distribution, and look for more such clouds around IC 10.

\begin{figure*}
\centering
   \subcaptionbox{\label{xig3:a}}{\includegraphics[width=2.5in]{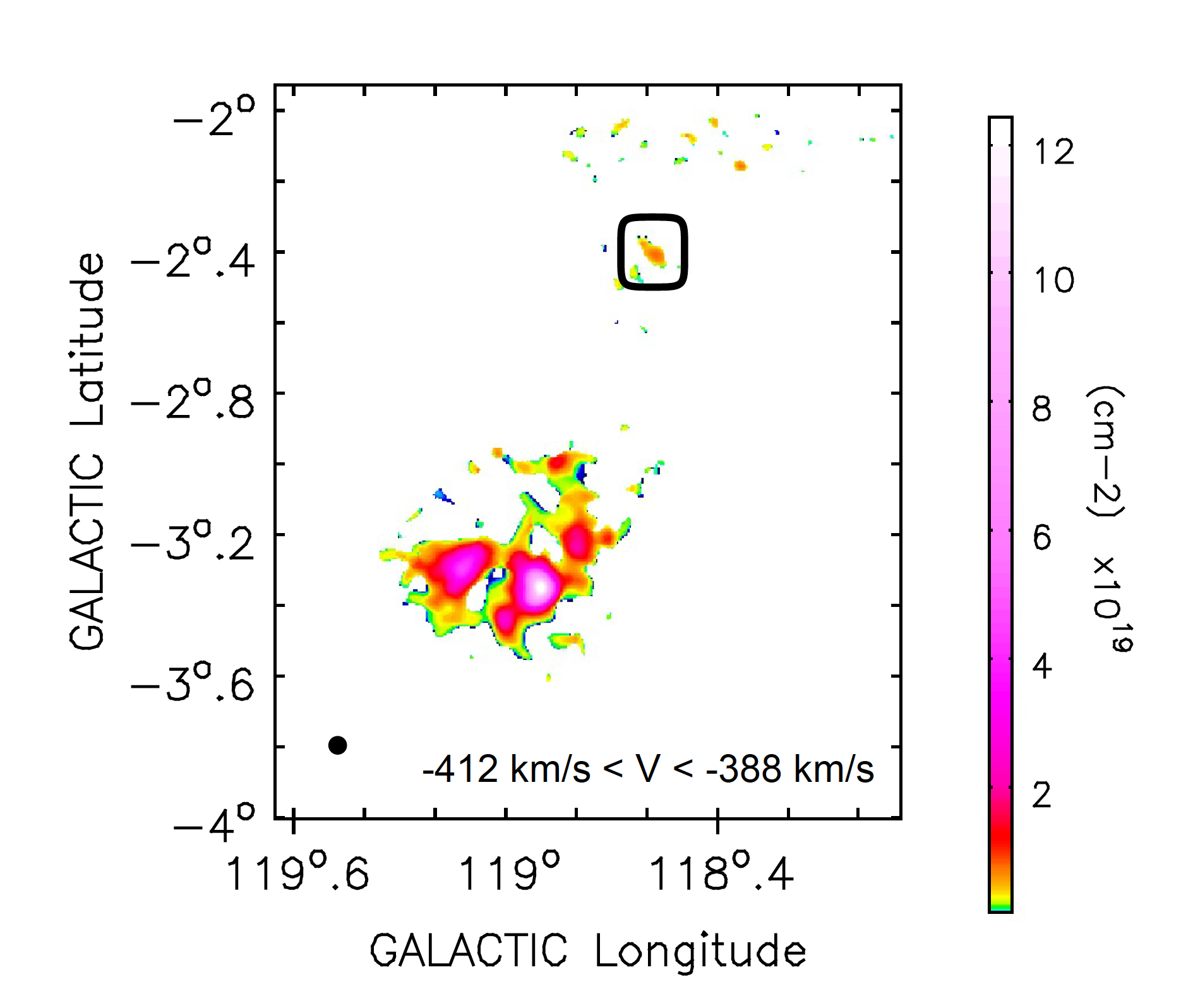}} \hspace{-1.2em} \\%
   \subcaptionbox{\label{xig3:b}}{\includegraphics[width=2.5 in]{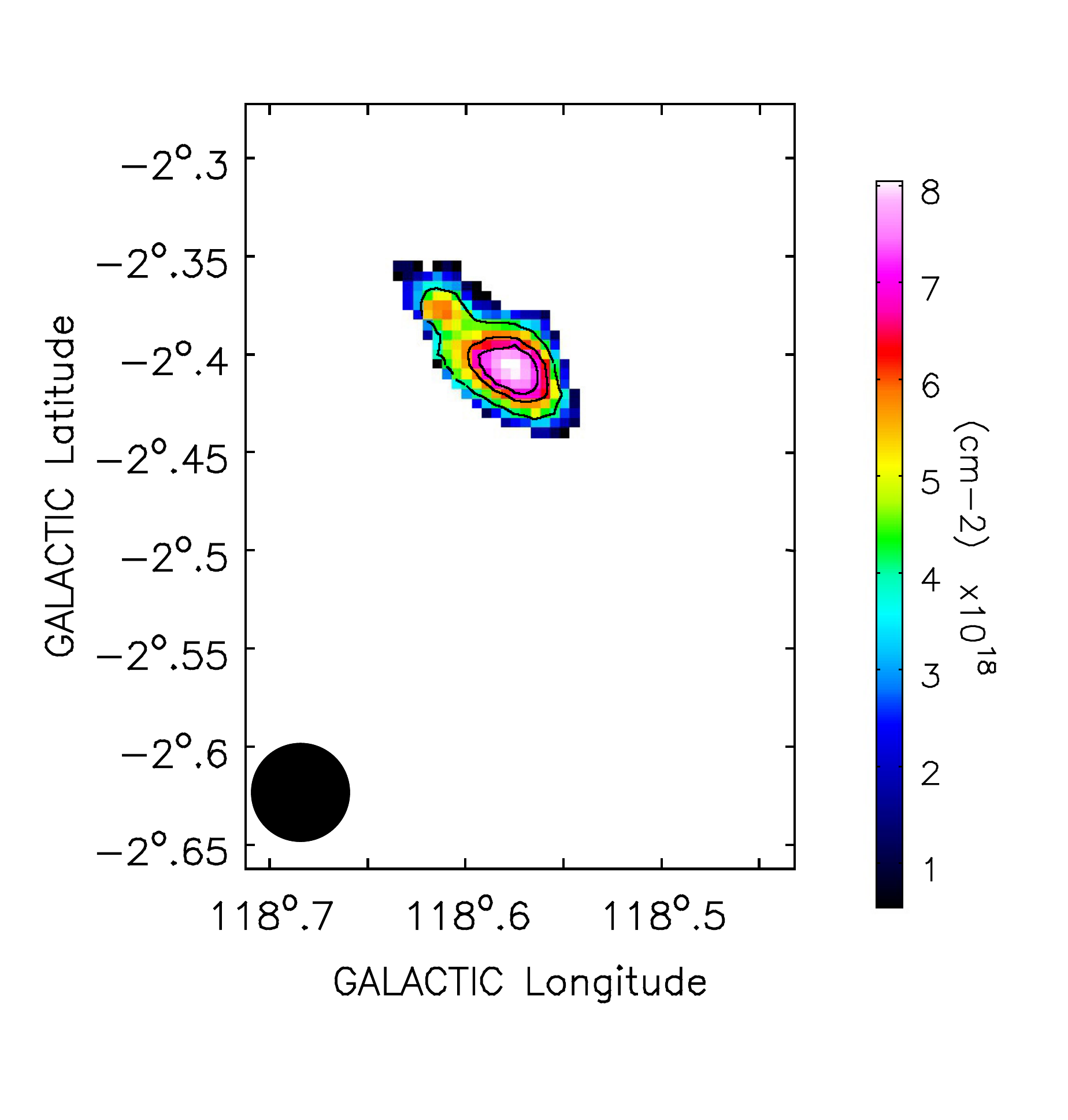}} \hspace{-1mm}
   \subcaptionbox{\label{xig3:c}}{\includegraphics[width=2.5in]{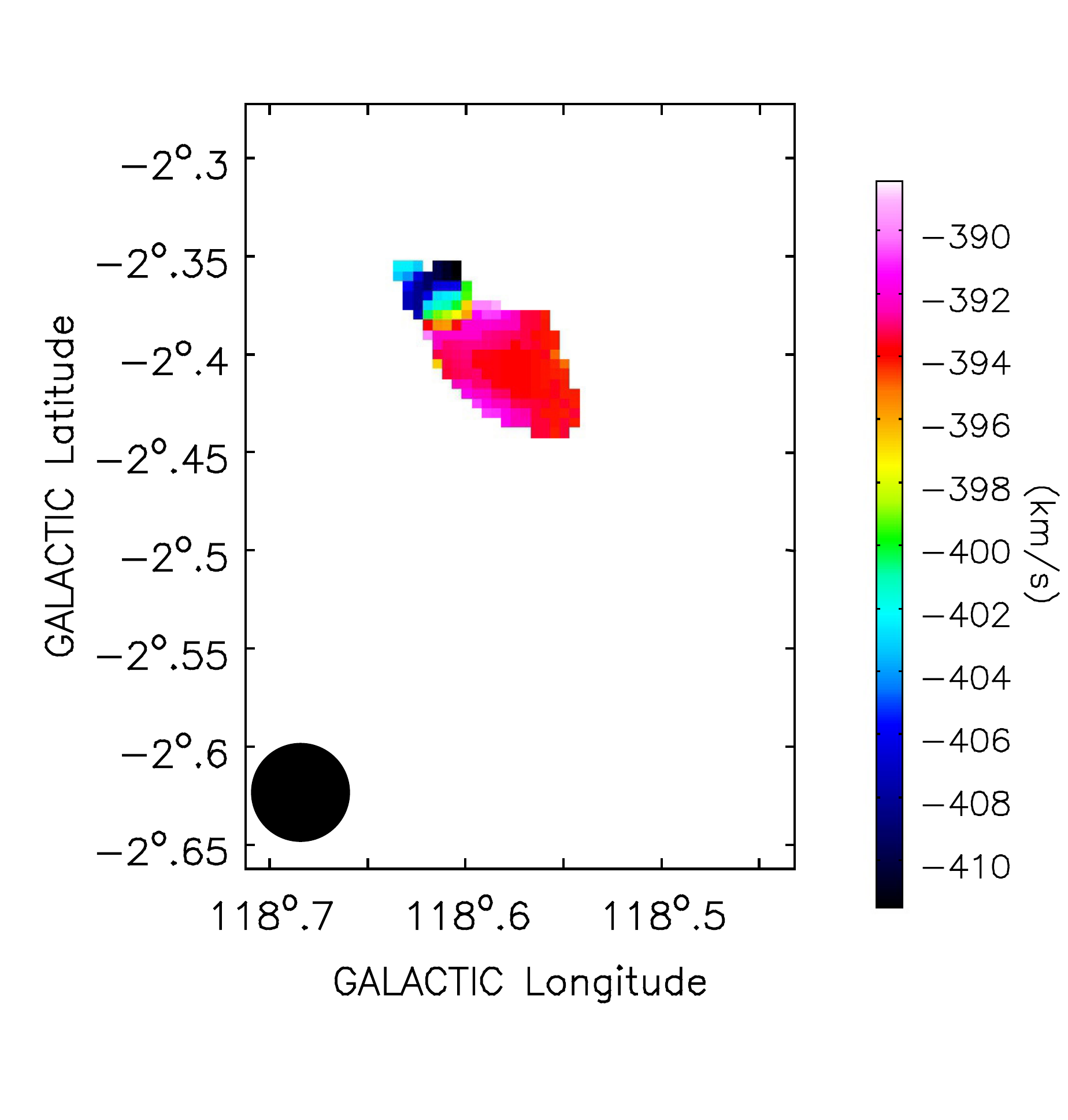}} \hspace{-1mm} \\
   \subcaptionbox{\label{xig3:d}}{\includegraphics[width=2.5in]{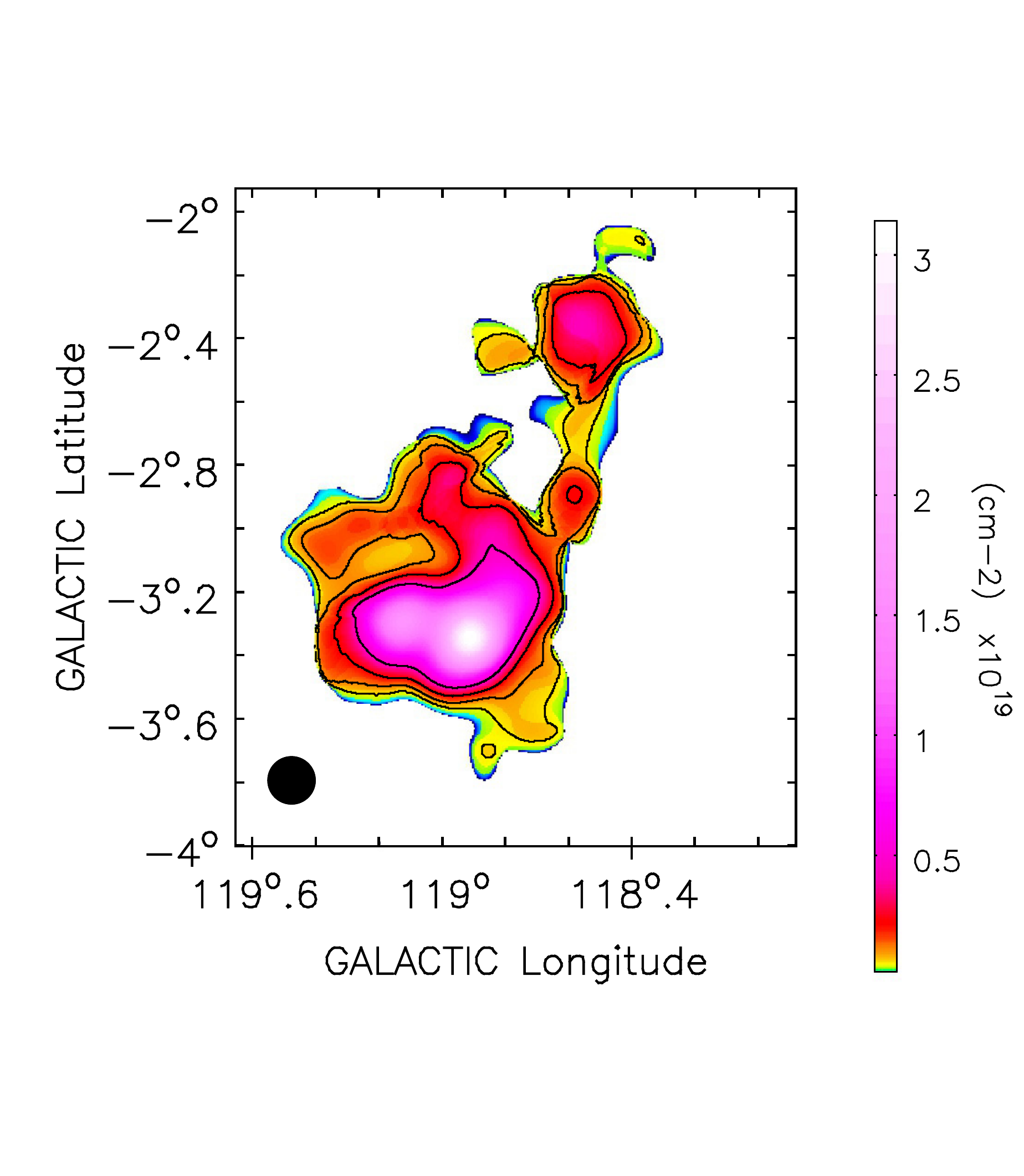}} \hspace{-1mm}
   \subcaptionbox{\label{xig3:e}}{\includegraphics[width=2.5in]{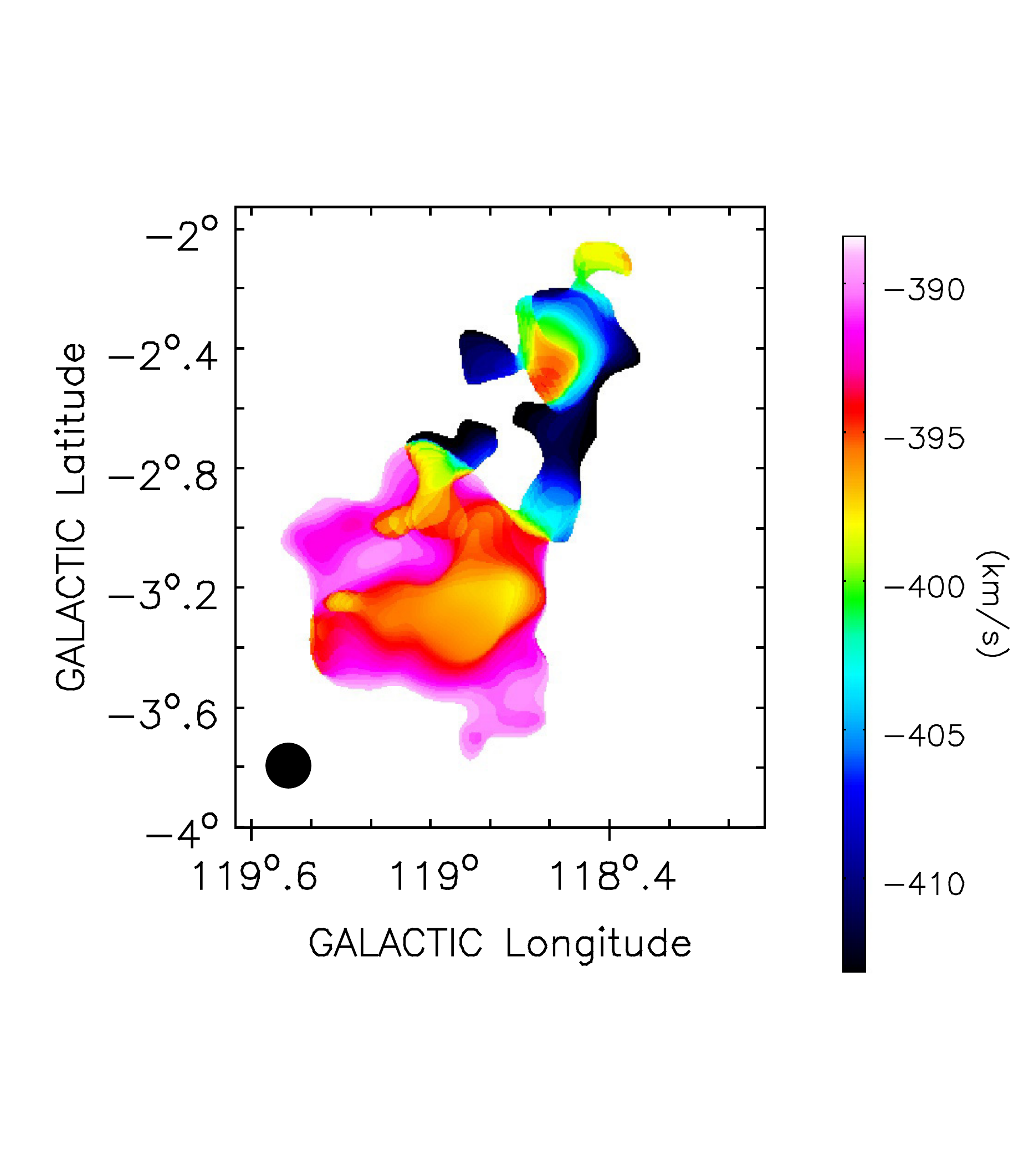}} \hspace{-1mm}    
   \caption{(a) Integrated column density map of the H\textsc{i} feature at 3$^{\prime}$ resolution for the velocity range -412 km\ s$^{-1}$ $\leq$ -388 km\ s$^{-1}$. The 3$\sigma$ column density limit at 3 arcmin is 4 $\times$ 10$^{18}$ cm$^{-2}$. At this column density, the H\textsc{i} cloud is clearly detected. (b) and (c) integrated column density and velocity field maps of the H\textsc{i} feature (-2$^{\circ}$.65 $\leq$ Galactic Latitude $\leq$ -2$^{\circ}$.3, 118$^{\circ}$.4 $\leq$ Galactic Longitude $\leq$ 118$^{\circ}$.7). The contours in (b) are 4, 6, 7 $\times$ 10$^{18}$ cm$^{-2}$.(d) and (e) integrated column density and velocity field maps of the H\textsc{i} feature at the GBT spatial resolution (9$^{\prime}$). The 3$\sigma$ column density limit at 9$^{\prime}$ is 8 $\times$ 10$^{17}$ cm$^{-2}$. The H\textsc{i} column density contours in (d) are at 8.2, 16.4, 32.8 and 65.6 $\times$ 10$^{17}$ cm$^{-2}$. A {}``bridge-companion'' structure between the H\textsc{i} cloud and IC 10 is visible at column densities $\sim$ 8 $\times$ 10$^{17}$ cm$^{-2}$. }
   \label{Fig:d} 
   \end{figure*}

Our DRAO observations are able to confirm the NW faint 21 cm H\textsc{i} emission feature discovered in the GBT observations. However, in order to be sure that the H\textsc{i} feature is real, our data had to be smoothed both in spatial and velocity resolution to increase the column density sensitivity. Figure \ref{xig3:a} shows the integrated column density map of IC 10 at 3$^{\prime}$ spatial resolution and 10.35 km\ s$^{-1}$ velocity resolution. The map was created by selecting the velocity range from -412 km\ s$^{-1}$ to -388 km\ s$^{-1}$. The 3$\sigma$ column density limit at 3$^{\prime}$ resolution is 4 $\times 10^{18}$ cm$^{-2}$. At this column density limit, we clearly resolve the Northern H\textsc{i} cloud seen in the GBT observations. Figure \ref{xig3:b} and \ref{xig3:c} shows the integrated column density and velocity field maps of the faint H\textsc{i} feature. The maps were created by selecting the region around the H\textsc{i} feature.

In order to examine our data at column densities comparable to the GBT limit, the DRAO data was smoothed to the GBT spatial resolution of 9$^{\prime}$. This allowed us to reach a 3$\sigma$ column density limit of 8 $\times$ 10$^{17}$ cm$^{-2}$ thus revealing the northern extension (bridge between the new H\textsc{i} cloud and IC 10, see Figure \ref{xig3:d} and \ref{xig3:e}) that was not visible at 3$^{\prime}$ resolution. The length and width  of the H\textsc{i} feature is found to be 0.9$^{\circ}$ and 0.3$^{\circ}$ respectively. Adopting a distance of 0.7 Mpc, an H\textsc{i} mass of 4.7 $\times 10^{5}$ M$_{\odot}$ is measured. 
 
As discussed in \citep{2013ApJ...779L..15N}, possible origins of the northern extension might include tidal or ram pressure, cold accretion, or stellar feedback to mention just a few. Looking at our data, the northern extension could either be a {}``bridge-companion'' structure or a tidal-tail structure as suggested by \citep{2014AJ....148..130A}). Although we are unable to investigate the origin of the H\textsc{i} feature in detail because the data had to be smoothed in spatial resolution to clearly confirm the H\textsc{i} cloud and bridge, the DRAO observations give a strong motivation for a need for deeper high spatial resolution H\textsc{i} observations on and around the H\textsc{i} feature that would allow us to investigate its origin and properties in more details.

\section{ Central H{\sc i} disk of IC 10} \label{s44}
%From the velocity field map in Figure 1(b), it is clear that the central region of IC 10 is a regularly rotating disk. In that sense, the rotation velocities of this region
From the velocity field map in Figure \ref{Fig:f}(a), it is clear that the central region of IC 10 is a regularly rotating disk. In that sense, the rotation velocities of this region can be determined. To separate the main disk from the rest of the galaxy, we first created regions in each channel around emission perceived to be associated with the disk (based on morphology and velocity). From this, a new cube consisting of only the selected emission was created. The MOMNT task in \textsc{aips} was used to derive the moment maps of the central part of IC 10. A second check in determining the extent of the central region was done using the dispersion map. For a regularly rotating disk, we expect the dispersion velocities to be high at the center and decrease uniformly with increasing radius. Using this approach, a mask was created around the velocity dispersion map excluding outer regions showing non-uniform dispersion values. This region was then applied to the intensity and velocity field maps. The final intensity, velocity field, and dispersion map of the central disk of IC 10 are shown in Figure \ref{Fig:f}(d), \ref{Fig:f}(a) and \ref{Fig:f}(e). A total flux of 196.2 $\pm$ 2.3 Jy km s$^{-1}$ is measured from the intensity map, corresponding to an H\textsc{i} mass of (2.300 $\pm$ 0.026) $\times$ 10$^{7}$ M$_{\odot}$ adopting a distance of 0.7 Mpc. This corresponds to $\sim$ 33$\%$ of the total H\textsc{i} mass of the complex. The dispersion velocities of the central region vary from $\sim$ 30 km s$^{-1}$ in the inner regions to $\sim$ 10 km s$^{-1}$ in the outer regions.

\begin{figure*}
\centering
   \includegraphics[width=3.55in]{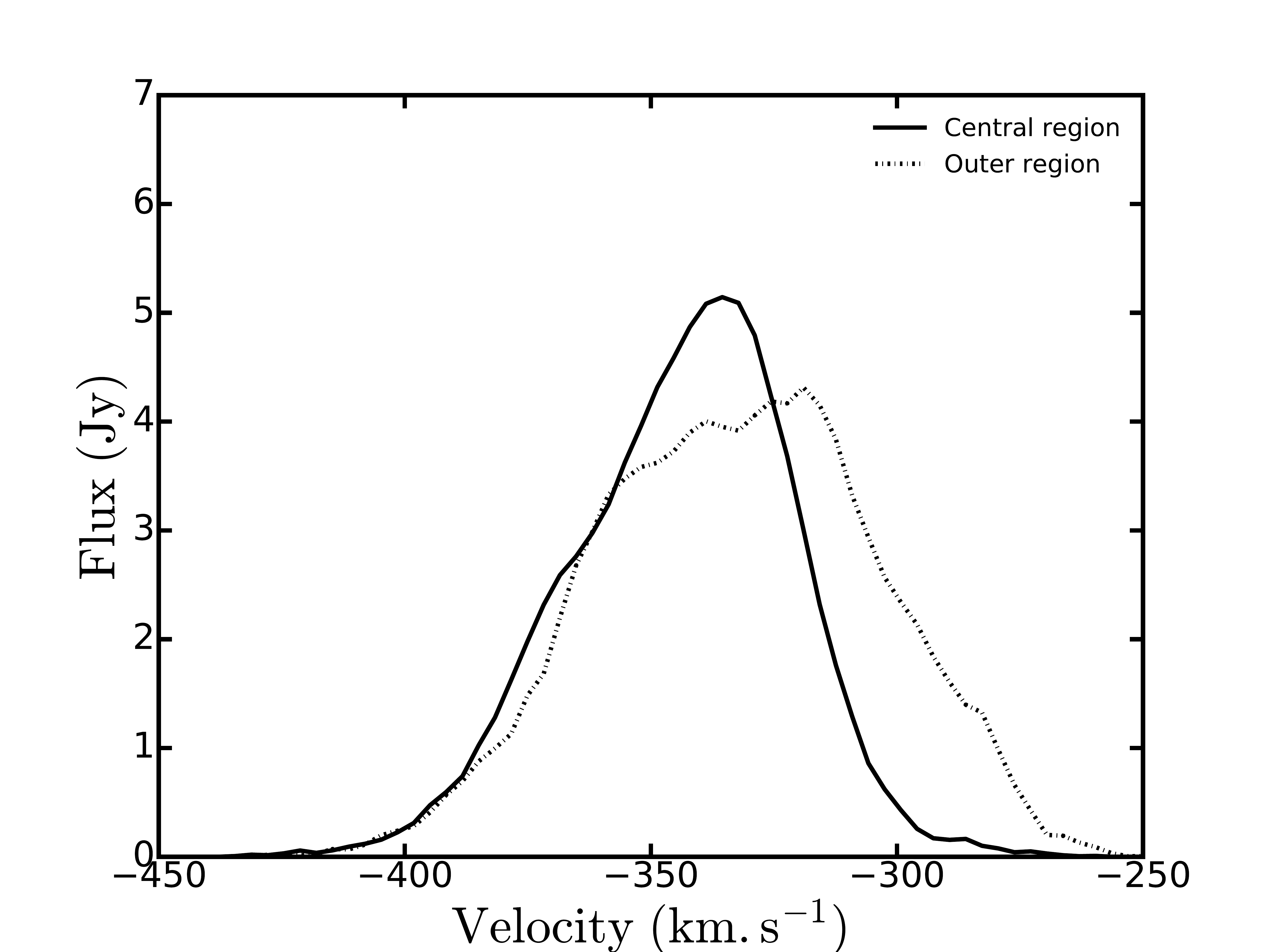} \hspace{-1.2em} %    
   \caption{Comparison of the H\textsc{i} global profile of IC 10 central region (solid line) and  the outer region of IC 10 (dash-dotted line).} 
   \label{Fig:dxx} 
   \end{figure*}
\subsection{Tilted-ring modeling}
We modeled the kinematics of the central part of IC 10 using the velocity field map shown in Figure \ref{Fig:f}(a). The rotation curve was derived by means of a tilted-ring model. 
The tilted ring model assumes that the gas rotates as a set of concentric rings, and that each ring can be described by the following dependent quantities: rotation velocity (V$_{\text{rot}}$), the systemic velocity (V$_{\text{sys}}$), the central position of the ring (x,y), the inclination ($i$), and the position angle (PA). The line of sight velocity at any position V(x,y) on a ring can be expressed as
\begin{equation}
V(x,y) = V_{\text{sys}} + V_{\text{rot}}\text{sin}(i)\text{cos}(\theta)
\end{equation}
where $\theta$ is an angle from the major axis of the galaxy plane and is related to the PA by the following relation:
\begin{equation}
\text{cos}(\theta) = \frac{-(x - x_{0}) \text{sin(PA)} + (y - y_{0})\text{cos(PA)}}{r},
\end{equation} 
\begin{equation}
\text{sin}(\theta) = \frac{-(x - x_{0}\text{cos(PA))} - (y - y_{0}\text{sin(PA))}}{\text{rcos(i)}}
\end{equation}
where r is the radius of the ring in the galaxy plane. The \textsc{gipsy} task ROTCUR \citep{1989A&A...223...47B} was used to fit tilted ring models to the H\textsc{i} velocity field along rings that are spaced by half the spatial resolution, i.e, 32$^{\prime \prime}$ in this case. A |cos$\theta$| weighting function and an exclusion angle of $\pm$ 10$^{\circ}$ about the minor axis were applied to the data points to give less weight to pixels close to the minor axis. First, we carried out ROTCUR with all initial parameters kept free (PA, $i$, x, y, V$_{\text{sys}}$, and V$_{\text{rot}}$). Assuming that the kinematical center and V$_{sys}$ do not vary from ring to ring, their mean values were determined from the first model and kept fixed throughout the fitting process. Different iterations were made either keeping PA or inclination or both fixed to determine the best fit model. After obtaining a satisfactory model for both sides, we made ROTCUR models of the approaching and receding sides separately.
 %the best fit model. After obtaining a satisfactory model for both sides, we made ROTCUR models of the approaching and receding sides separately.

\subsection{Results of the tilted ring fit}
\begin{figure}
\centering
  \includegraphics[width=3.0in]{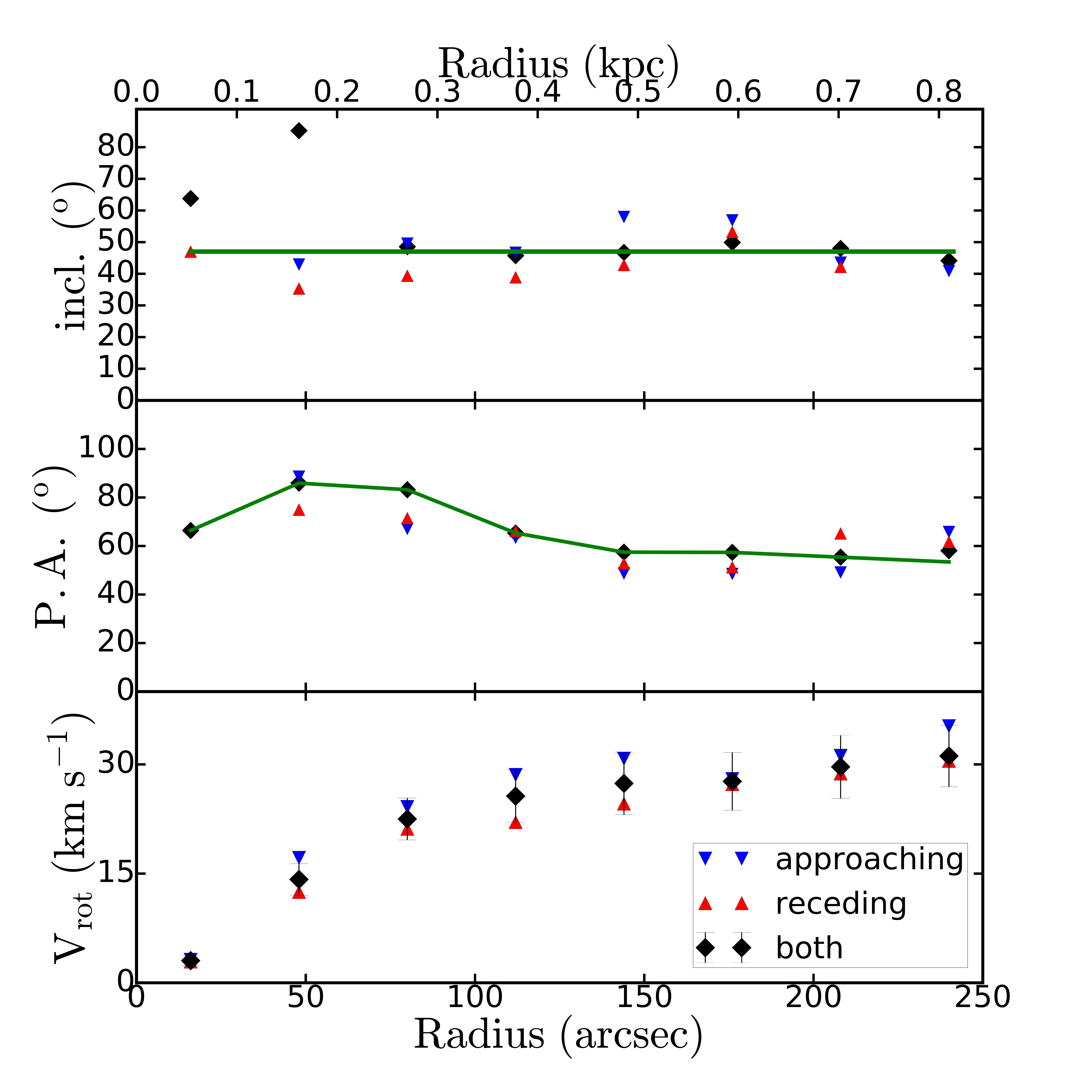}\hspace{-1mm}\\%    
   \caption{Results of the tilted ring fit of IC 10. The top panel shows the inclination, the middle panel the major axis position angle, and the bottom panel shows the final rotation curves. Red upward triangles are the results of the receding side, blue downwards triangles for the approaching side, and the black diamonds represent both sides. The green solid lines are the adopted values used to derive the final rotation velocities.}
   \label{Fig:e}  
   \end{figure}

\begin{figure*}
\centering
   \subcaptionbox{Observed velocity field\label{fig3:a}}{\includegraphics[width=2.8in]{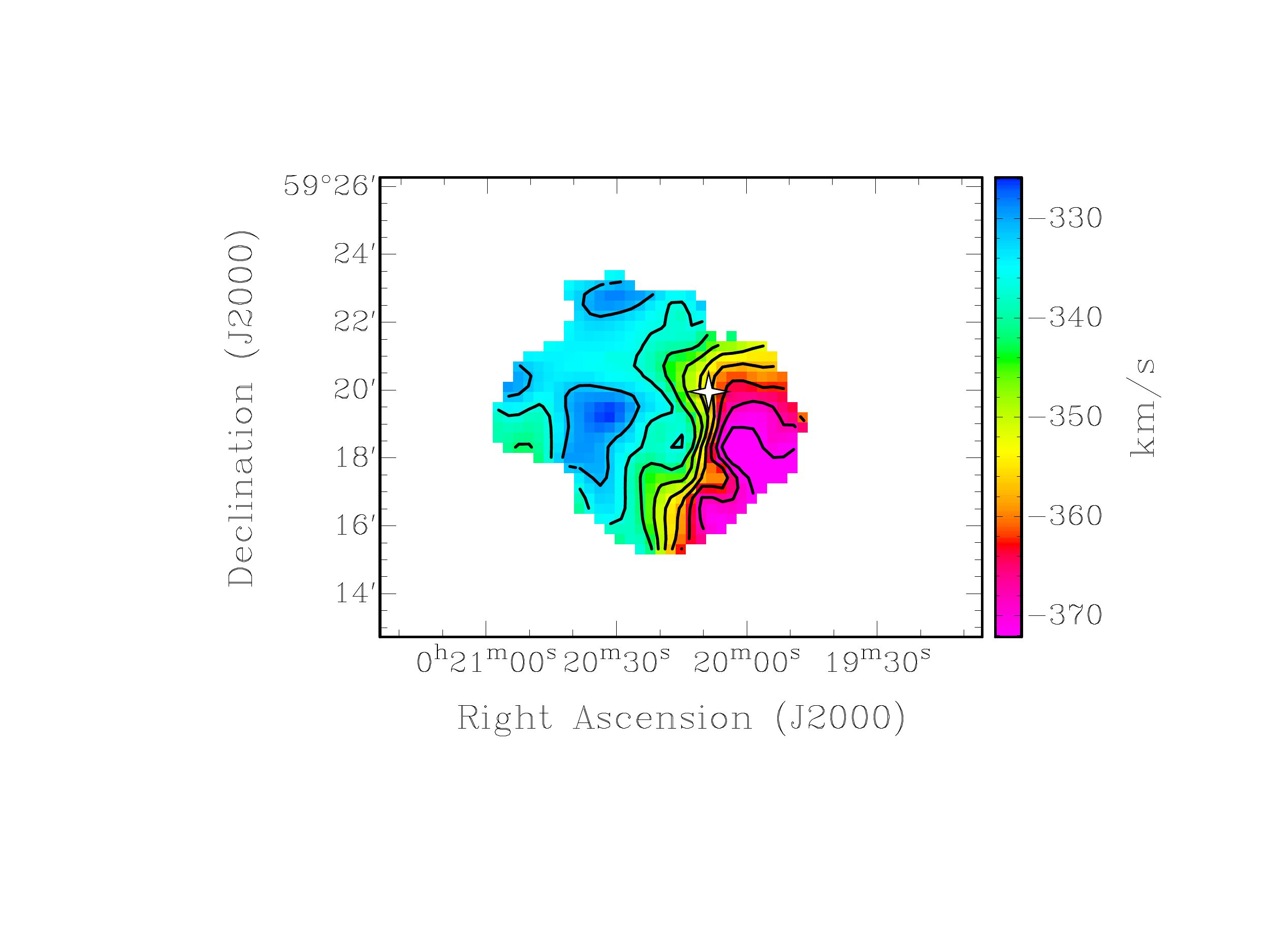}} \\ %
   \subcaptionbox{Model velocity field map \label{fig3:b}}{\includegraphics[width=2.5in]{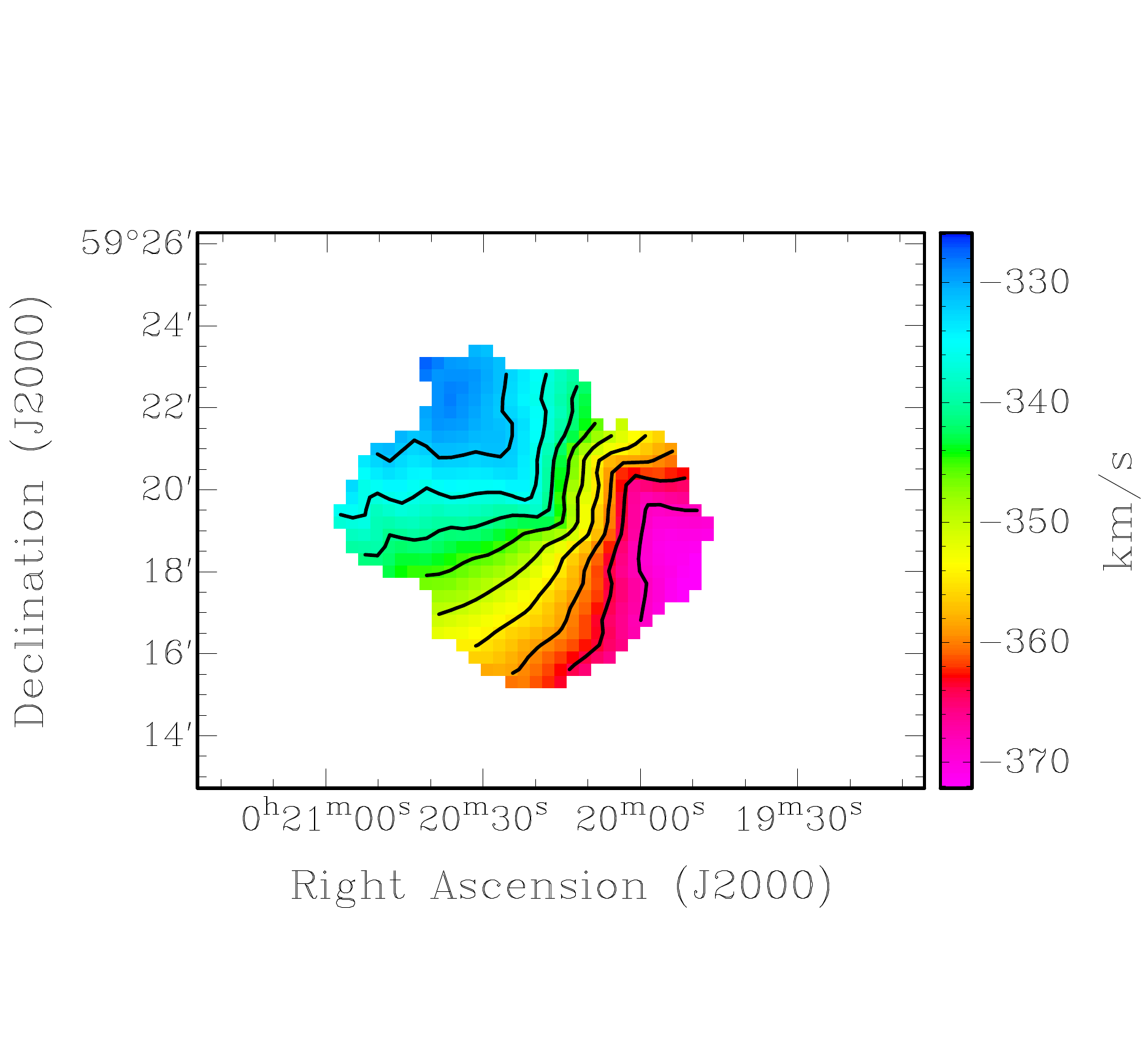}} \hspace{-1mm}  
   \subcaptionbox{Residual map \label{fig3:a}}{\includegraphics[width=2.5in]{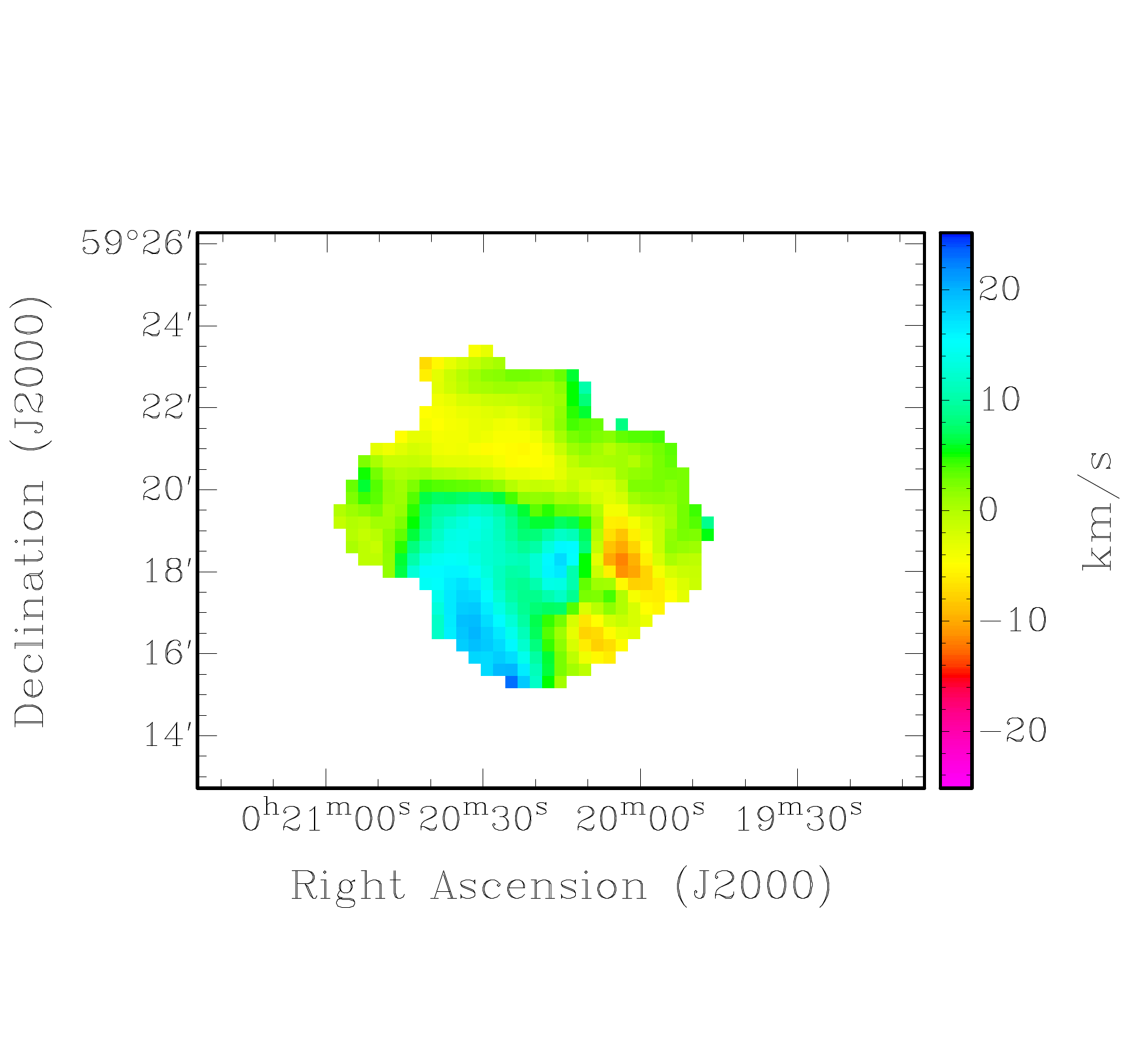}}\hspace{-1mm}\\
   \subcaptionbox{Intensity map\label{fig3:b}}{\includegraphics[width=2.5in]{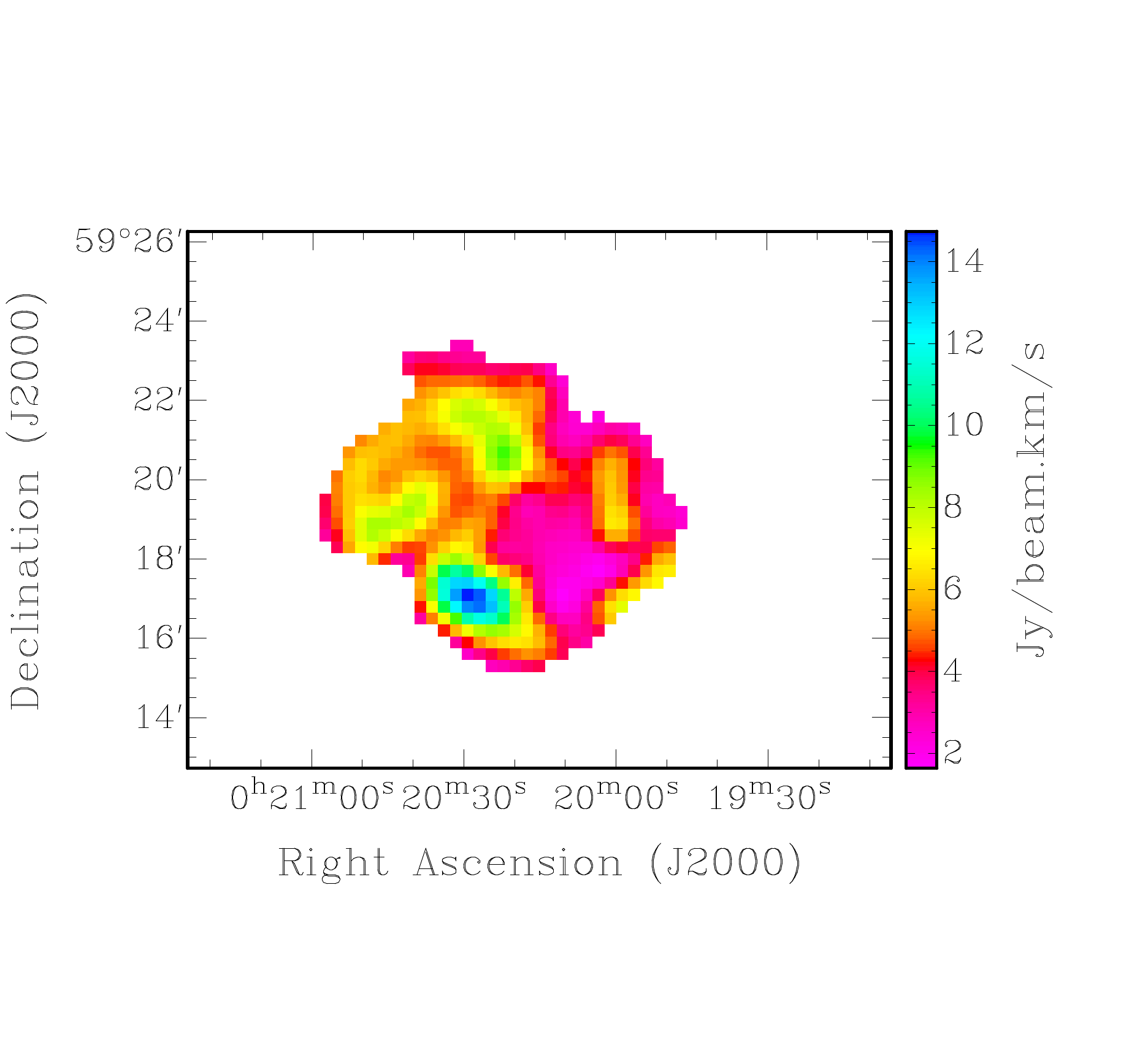}} \hspace{-1mm} 
   \subcaptionbox{Velocity dispersion map\label{fig3:a}}{\includegraphics[width=2.5in]{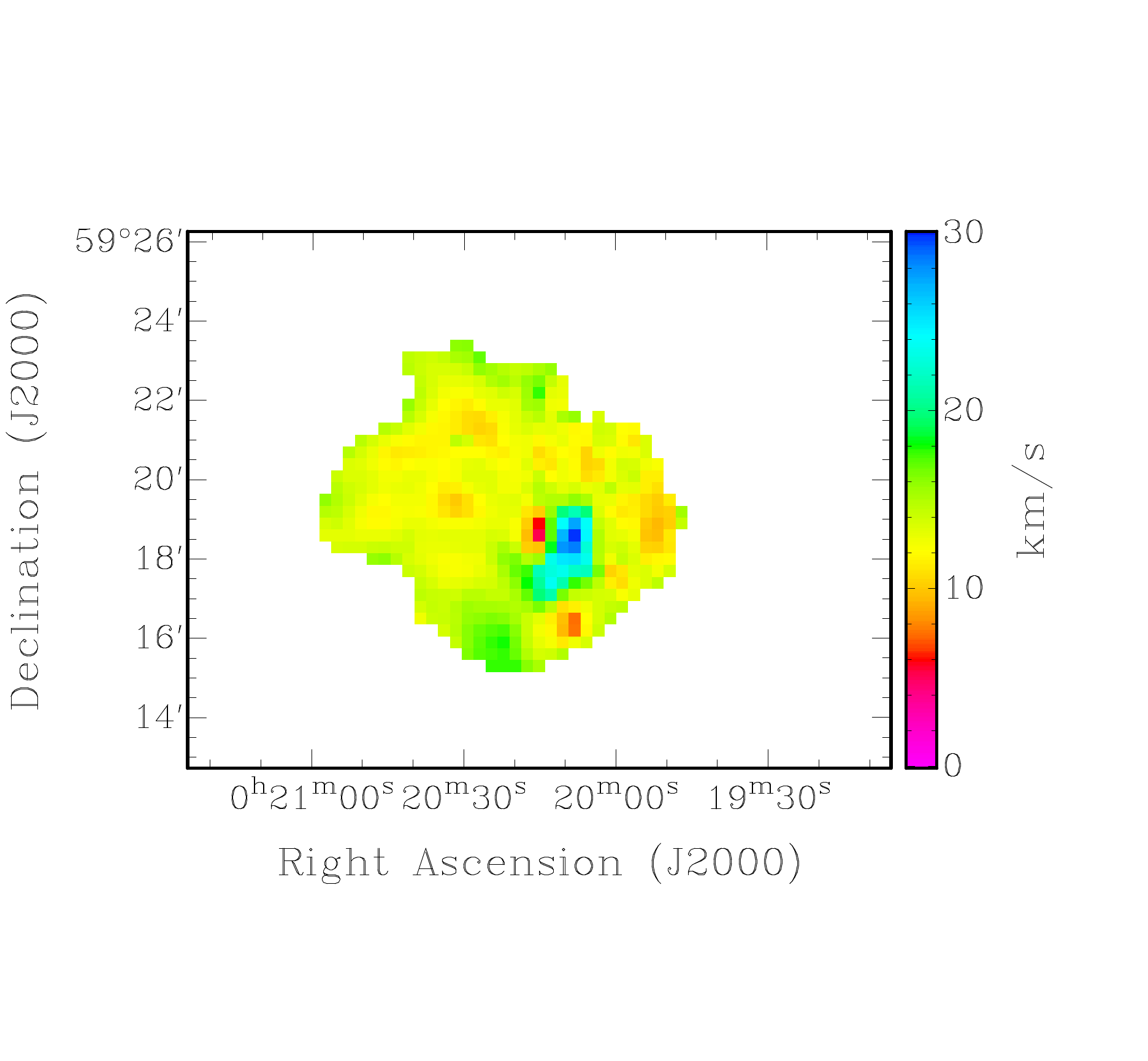}}\hspace{-1mm}\\%    
   \caption{Maps of the central disk of IC 10: Observed velocity field \ref{Fig:f}(a), model velocity field map \ref{Fig:f}(b), residual map \ref{Fig:f}(c), intensity map \ref{Fig:f}(d), and velocity dispersion map \ref{Fig:f}(e). The observed and model velocity field contours run from -370 to -330 km s$^{-1}$ in steps of 5 km s$^{-1}$. The white star in Figure \ref{Fig:f}(a) represents the position of the kinematical center.}  
   \label{Fig:f} 
   \end{figure*}

The results of the tilted ring model to the velocity field are shown in Figure \ref{Fig:e} and \ref{Fig:f}. The resulting kinematical parameters are presented in Table \ref{tab:template3}. The H\textsc{i} rotation curve of the central disk of IC 10 is measured out to 0.81 kpc. The rotation curve of IC 10 is similar in form to those of other dwarf galaxies \citep{10.1093/mnras/stx2256}, having a linearly rising (solid body) inner portion and then turning over to become approximately flat. On the other hand, the slope of the inner part of IC 10 is $\sim$ 3 times steeper (65 km\ s$^{-1}$/kpc) than what has been derived for most dwarf galaxies. \citet{10.1093/mnras/stx2256,10.1093/mnras/sty1056} calculates the inner slopes of 21, 21, and 13 km\ s$^{-1}$/kpc for NGC 6822, Sextans A and B while \citet{1996AJ....111.1551M} and \citet{1998MNRAS.300..705M} derive the inner slopes of 25 km\ s$^{-1}$/kpc and 22 km\ s$^{-1}$/kpc for BCD galaxies NGC 2915 and NGC 1705. We find the kinematical V$_{sys}$ = -351 $\pm$ 1.8 km s$^{-1}$, this value is different from the V$_{sys}$ of -339 $\pm$ 6 km s$^{-1}$ derived from the global profile due to the asymmetry. A mean PA = 65$^{\circ}$ $\pm$ 4$^{\circ}$ and $i$ = 47$^{\circ}$ $\pm$ 6$^{\circ}$ are found. These values are in agreement with the results of \citet{2015AJ....149..180O} who measured V$_{sys}$ = -348 km s$^{-1}$, PA $\sim$ 60$^{\circ}$, and $i$ $\sim$ 48$^{\circ}$. \citet{1989A&A...214...33S} measured a V$_{sys}$ = -352 km s$^{-1}$ and slightly lower values of $i$ = 40$^{\circ}$ and PA = 50$^{\circ}$. Figure \ref{Fig:g} compares the IC 10 DRAO rotation curve with the rotation curve derived by \citet{2015AJ....149..180O} from high resolution VLA LITTLE THINGS data. The two curves agree well within the errors.

\begin{figure}
  % Maximum length
  \includegraphics[width=2.8in]{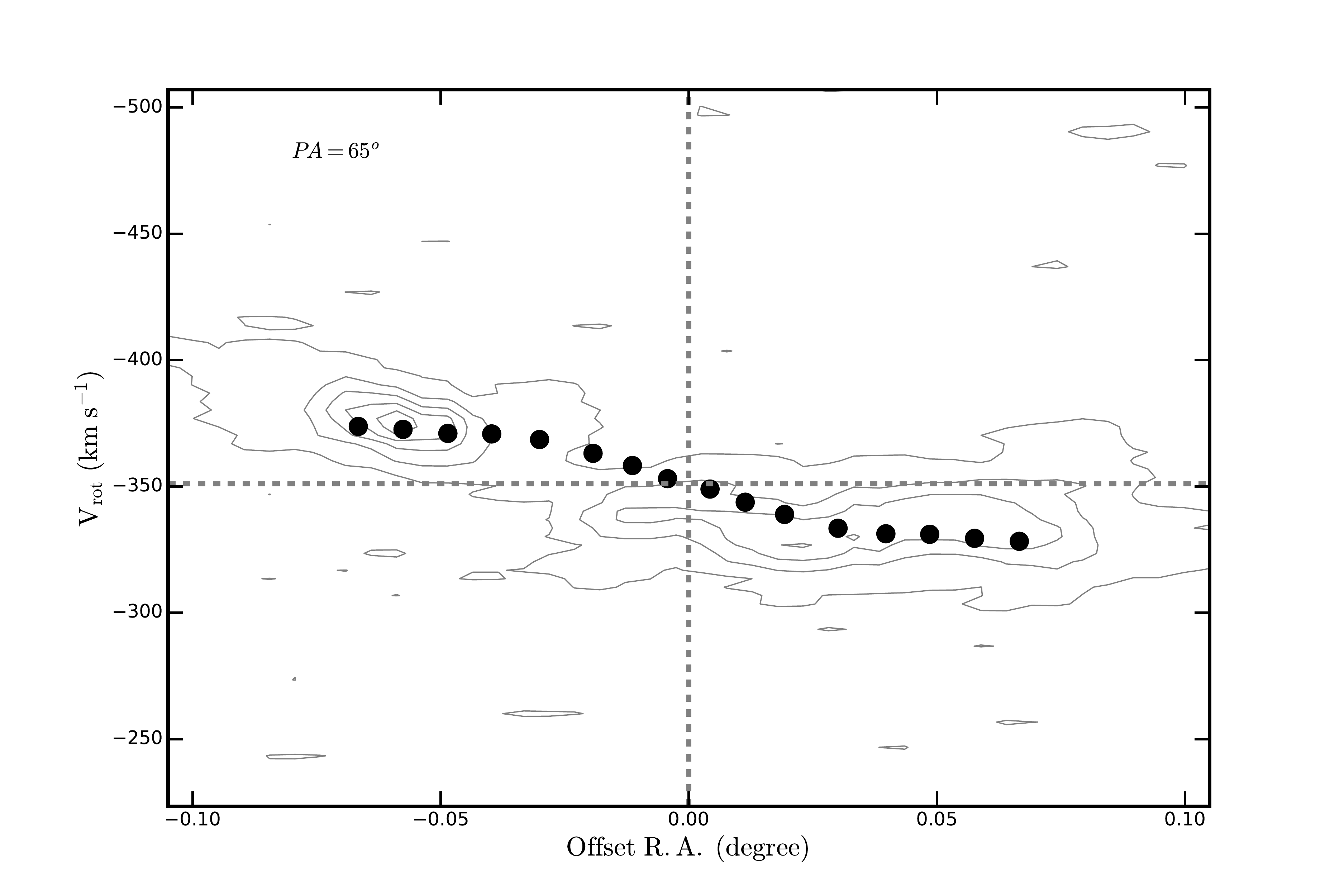}\hspace{-1.2em}%
   \vspace*{6mm}
   \caption{Position velocity diagram of IC 10 central regions with the projected rotation curve over plotted. The dotted grey lines represent the centre of the galaxy and the systemic velocity. Superimposed is the rotation curve derived from the tilted ring model, corrected for the inclination of a slice along the galaxy major axis.} 
   \label{Fig:2019pv} 
\end{figure}

\begin{figure}
  \includegraphics[width=3.0in]{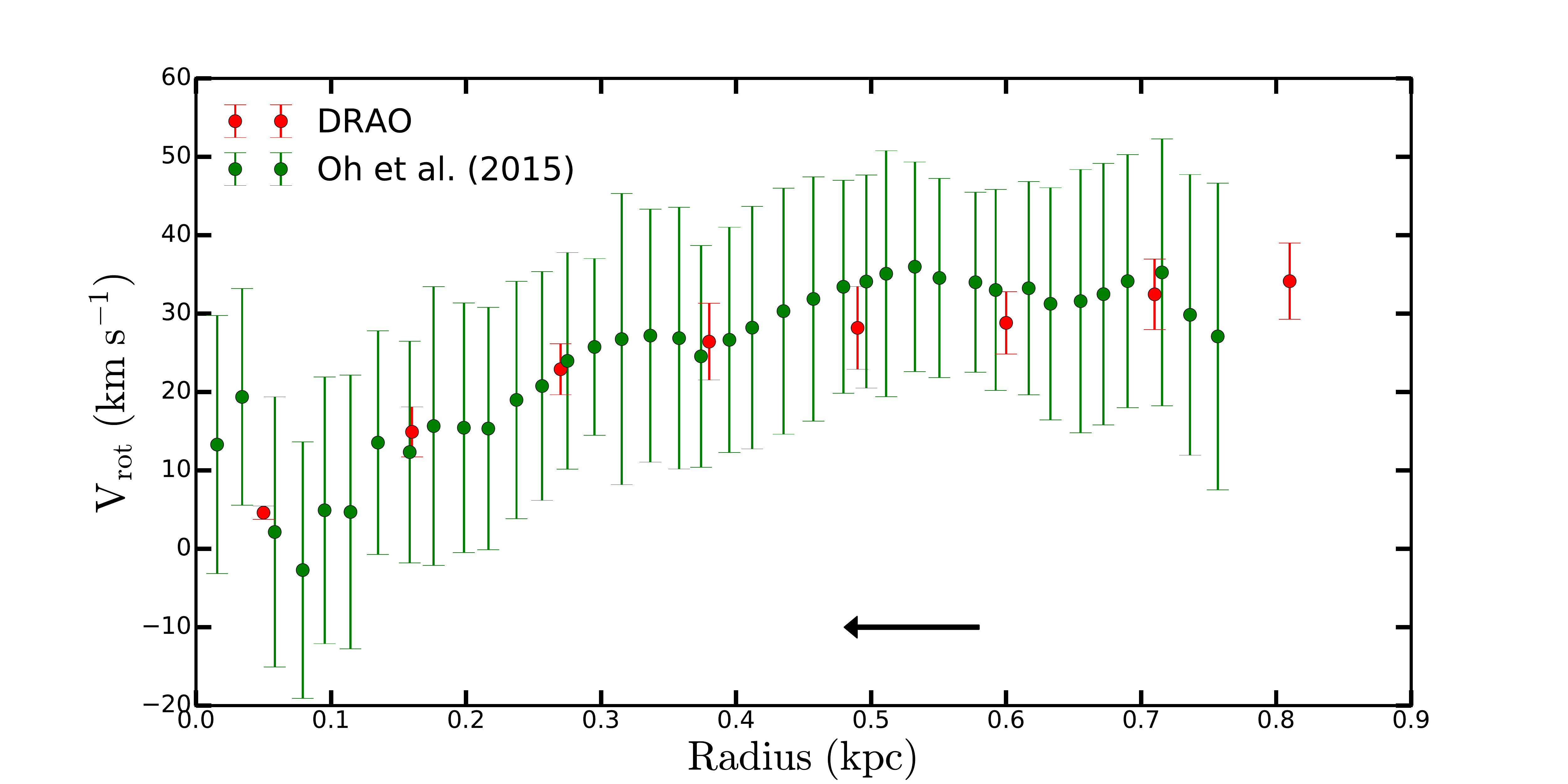}\hspace{-1mm}\\%    
   \caption{Comparison of the IC 10 DRAO rotation curve (red circles) with the analysis by \citet{2015AJ....149..180O} (green circles).  The black arrow indicates the region used in the \citet{2015AJ....149..180O} analysis.} 
   \label{Fig:g} 
   \end{figure}

To check the reliability of our derived kinematical parameters, we built a model velocity field (\ref{Fig:f}(b)) using the best-fitting values and compared this to the observed velocity field (Figure \ref{Fig:f}(a)). The two maps agree well with small deviations. The residual map (Figure \ref{Fig:f}(c)), which is the difference between the observed and the model velocity field does not show any large scale deviations which implies that the derived velocity field model is in good agreement with the observed velocity field. The mean value of the residual map is 1.0 km s$^{-1}$ with a rms spread of 6.2 km s$^{-1}$. The residual velocity field shows some large scale structure in the sense that the south-east part shows negative residuals and the north and west parts predominantly positive residuals which are consistent with the velocity field in Figure \ref{Fig:f}. This indicates that the inner disk could be associated with non circular motions. Figure \ref{Fig:2019pv} shows the rotation curve overlaid on a position-velocity (PV) diagram obtained along the major axis. 

The errors on the final rotation curve are calculated as the quadratic sum of the dispersion in each ring and half the difference between the approaching and receding sides:
\begin{equation}
\Delta V = \sqrt{\sigma^{2} + \bigg(\frac{|V_{app} - V_{rec}|}{2}\bigg)^2}
\end{equation}
The rotation velocities were corrected for the asymmetric drift. Following the procedure used by \citet{2000AJ....120.3027C}, the corrected rotation velocities are given by:
\begin{equation}
V^{2}_{c} = V_{rot}^{2} - 2\sigma \frac{\delta \sigma}{\delta \ln R} - \sigma^{2} \frac{\delta \ln \Sigma}{\delta \ln R}
\end{equation}


 \begin{table}   
%\scriptsize
\caption{\small Radial H\textsc{i} distribution and kinematical parameters of IC 10.}
\begin{minipage}{\textwidth}
\begin{tabular}{l@{\hspace{0.10cm}}c@{\hspace{0.10cm}}c@{\hspace{0.10cm}}c@{\hspace{0.10cm}}c@{\hspace{0.08cm}}c@{\hspace{0.08cm}}c@{\hspace{0.05cm}}}   
\hline

Radius & $\Sigma_{g}$   &  $\sigma $ &  V$_{\text{rot}}$ & $\Delta$ V & V$_{\text{corr}}$ &   V$_{\text{{c}}}$  \\
arcsec & M$_{\odot}$pc$^{-2}$    &   km s$^{-1}$   &    km s$^{-1}$ & km s$^{-1}$ & km s$^{-1}$ & km s$^{-1}$ \\ \hline \hline
0.0&9.4&13.7&0.0&0.0&0.0&0.0\\
16&9.1&13.2& 3.0&0.7&1.8&4.9\\
48&9.3&12.8&14.2&3.2&0.6&14.8\\
80&9.9&12.4&22.5&3.2&0.3&22.8\\
112&9.7&12.5&25.6&4.8&0.8&26.4\\
144&9.0&12.7&27.4&5.3&0.7&28.1\\
176&8.8&12.4&27.6&3.9&2.1&29.7\\
208&9.7&13.3&29.7&4.5&3.8&33.5\\
240&10.6&13.2&31.1&4.8&2.3&33.4\\

\hline    
\multicolumn{7}{@{} p{8 cm} @{}}{\footnotesize{\textbf{Notes.} Column (1) gives the radius, column (2) the surface densities, column (3) the velocity dispersion, column (4) the observed rotation velocities, column (5) the errors of those velocities, column (6) correction from asymmetric drift, and column (7) the corrected velocities used for the mass models.}}
\label{coords_table}
%\end{center} 
\end{tabular}  

\end{minipage}
\label{tab:template3}
\end{table}

\subsection{Mass modeling}
The rotation velocity is a direct reflection of the gravitational potential of the galaxy, and can therefore be used to derive the total distribution of matter. The gravitational potential is the sum of the individual mass components in each galaxy. Expressed in velocities, this sum becomes:
\begin{equation}
V_{\text{rot}}^{2} = V_{g}^{2} + V_{d}^{2} + V_{h}^{2}
\end{equation}
with V$_{\text{rot}}$ being the observed total rotation velocity, V$_{g}$ the contribution of the gas disk to the total rotation curve,  V$_{d}$ the contribution of the stellar disk and V$_{h}$ the contribution of the dark matter halo component.
\subsubsection{Stellar and gas component}
\begin{figure}
  % Maximum length
  \centering
  \includegraphics[width=3.8in]{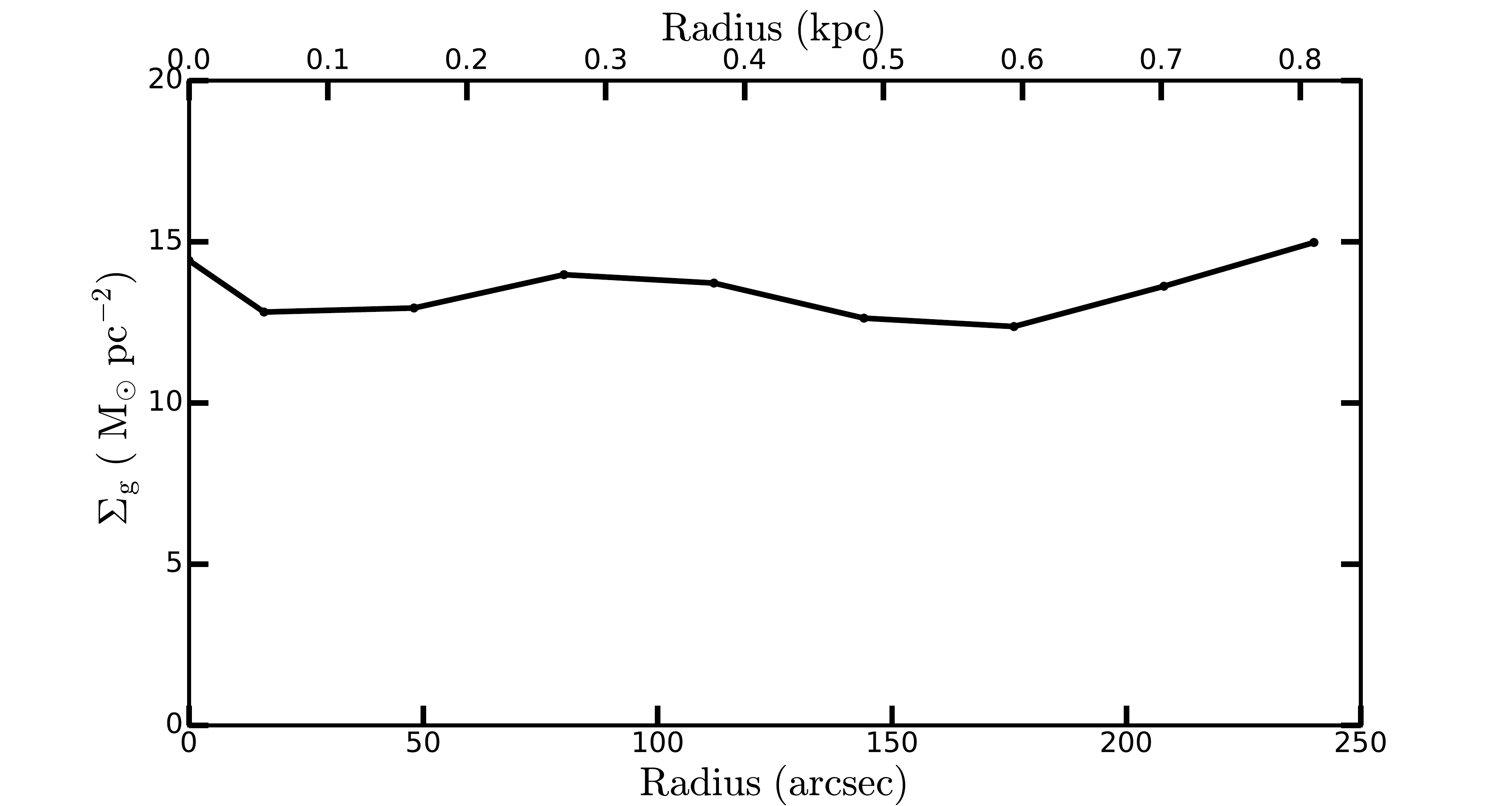}\hspace{-1.2em}%
   \caption{H\textsc{i} surface density profile of IC 10 derived from \textsc{gipsy} task ELLINT.} 
   \label{Fig:h} 
\end{figure}

The contribution of the gaseous disk was computed using the H\textsc{i} surface density profile shown in Figure \ref{Fig:h}. The profile was derived from the H\textsc{i} column density map using the \textsc{gipsy} task ELLINT and applying the kinematical parameters derived from the tilted ring fit. The H\textsc{i} surface distribution was then multiplied by a factor of 1.4 to account for helium and other metals. The gas surface density profile were then converted to the corresponding gas rotation velocities assuming the gas component is mainly distributed in a thin disk. This was done using the \textsc{gipsy} task ROTMOD. We used the WISE 3.4$\mu$m surface brightness profile to characterize the stellar distribution in IC 10. The 3.4$\mu$m WISE band traces light from the old stellar population and therefore is an effective measure of stellar mass. The surface brightness profile in mag arcsec$^{-2}$ was converted into luminosity density profile in units of L$_{\odot}$ pc$^{-2}$ and then to mass density using the expression
\begin{equation}
\Sigma[M_{\odot} pc^{-2}] =  (M/L)_{*,3.4} \times 10^{-0.4 \times (\mu_{3.4} - C_{3.4})}
\end{equation}
where $\mu_{3.4}$ is the stellar brightness profile, (M/L)$_{*,3.4}$ is the mass to light ratio in the 3.4$\mu$m WISE band, and C$_{3.4}$ is the constant used for conversion from mag arcsec$^{-2}$ to L$_{\odot}$ pc$^{-2}$ and is calculated as C$_{3.4}$ = M$_{\odot}$ + 21.56. M$_{\odot}$ = 3.24, is the absolute magnitude of the Sun in the 3.4$\mu$m WISE band. Using the stellar mass densities, the GIPSY task \textsc{ROTMOD} was used to compute the corresponding stellar velocities. The mass to light ratio was calculated based on the stellar distribution models using the (W1-W2) color. Using the method of \citet{2014ApJ...782...90C}, the M/L$_{3.4}$ at 3.4$\mu$m is given by
\begin{equation} \label{eq1}
\log_{10}M_{\text{stellar}}/L_{\text{W1}} = -1.93(W_{3.4\mu m} - W_{4.6 \mu m}) - 0.04
\end{equation}
with (W1-W2) = 0.08, a mass to light ratio of 0.6 is derived for IC 10. However, Equation \ref{eq1} may not be usable for certain dwarf galaxies with bluer colors. 

\subsubsection{Dark matter models}
We have considered two different models to study the properties of dark matter in IC 10: the Navarro Frenk and White (NFW) \citep{1997ApJ...490..493N} (NFW) and the pseudo-isothermal \citep{1991MNRAS.249..523B} (ISO) models. The NFW profile is derived from the results of dark matter only cosmological simulations performed within the frame of the $\Lambda$CDM theory. The density $\rho_{\text{NFW}}$ is given by
\begin{equation}
\rho_{\text{NFW}}(r) = \frac{\rho_{s}}{\bigg(r/r_{s}\bigg) \bigg(1 + r/r_{s}\bigg)^2},
\end{equation}
where $\rho_{s}$ and $r_{s}$ are the characteristic density and scale radius of the NFW halo. The NFW halo rotation curve is given by
\begin{equation}
V(r) = V_{200}\bigg[\frac{\ln(1 + cx) - cx/(1 + cx)}{x[\ln(1 + c) - c/(1 + c)}\bigg]^{1/2}
\end{equation}
where c = r$_{200}$/r$_{s}$ is the concentration parameter, x = r/r$_{200}$, and V$_{200}$ is the characteristic velocity. The radius r$_{200}$ is the radius where the density contrast with respect to the critical density of the universe exceeds 200, roughly the virial radius. \\

The ISO model results from the shape of the observed rotation curves, and models the halo as a central constant-density core. The form of this core-like halo is given by:
\begin{equation}
\rho_{\text{ISO}}(r) = \frac{\rho_{0}}{1 + \bigg(r/r_{c}\bigg)^2}
\end{equation}
where $\rho_{0}$ is the central density and r$_{c}$ is the scaling radius. The corresponding halo rotation curve is given by
\begin{equation}
V(r) = 4\pi G \rho_{0} r_{c}^{2}\bigg[1 -\frac{r_{c}}{r} \text{arctan}\bigg (\frac{r}{r_{c}}\bigg) \bigg]
\end{equation}

\subsubsection{Mass modeling results} \label{mass}
\begin{figure}
  % Maximum length

\includegraphics[width=3.0in]{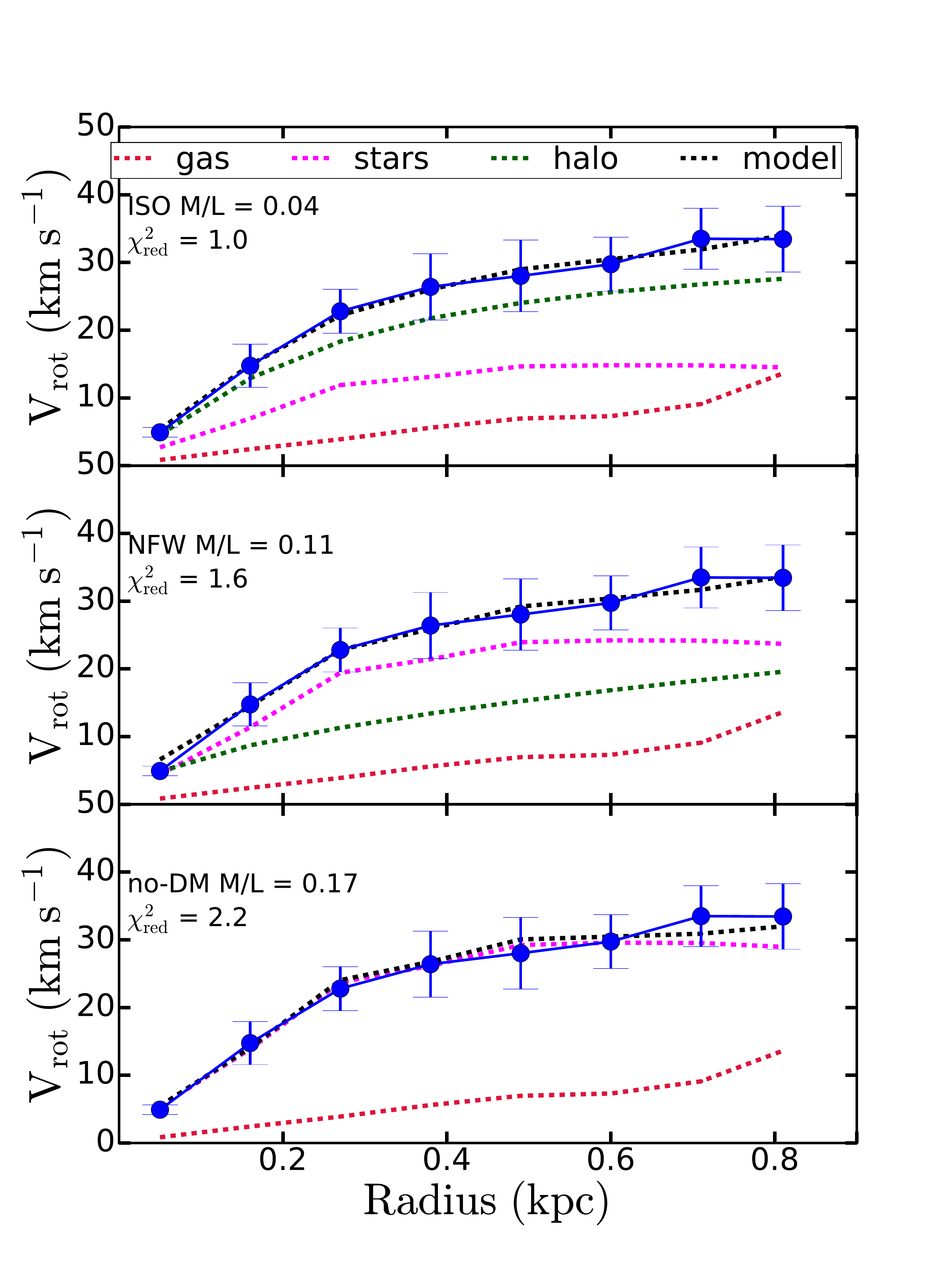}\hspace{-1.2em}%
   \caption{Mass distribution models of IC 10 with ISO (top panel), NFW (middle panel) and no dark matter halo (bottom panel). The blue filled circles present the observed rotation curve, the black dotted lines the model rotation curve, the darkgreen dotted lines indicate the dark matter rotation velocities, the crimson dotted lines show the gas rotation velocities, and the magenta dotted lines show the rotation velocities of the stellar components.}  
   \label{Fig:i}
\end{figure}

The results of the mass modeling of IC 10 are summarized in Figure \ref{Fig:i} and Table \ref{tab:template4}. The \textsc{gipsy} task ROTMAS was used to decompose the observed rotation curve into luminous and dark matter components. The inverse squared weighting of the rotation curve data points with their uncertainties were used during the fitting. The M/L value of 0.6 derived from Equation \ref{eq1} was too large to fit the stellar component of IC 10, therefore we used the best fitting model by letting M/L freely vary. It can be seen from Figure \ref{Fig:i} that the ISO model reproduces better the observed rotation curve of the inner disk of IC 10 with a reduced $\chi^{2}$ value of 1.0. However, a M/L of 0.04 ($\sim$ 15 times smaller than predicted from the WISE infrared color) derived from this fit is too small to make physical sense. On the other hand, the NFW fit gives a reduced $\chi^{2}$ value of 1.6. The difference between the two models is seen in the inner most part where the ISO model fits better than the NFW model. The large errors derived for the NFW fitted parameters (see Table \ref{tab:template4}) suggest that this model does not produce a good fit to the observed rotation curve. Moreover, the value of the concentration parameter c found is unphysical: c = 1 means no collapse and c < 1 is impossible in the $\Lambda$CDM context \citep{2008AJ....136.2648D}. The bottom panel of Figure \ref{Fig:i} shows the results of the mass model fit without a dark matter component. Although a larger $\chi^{2}$ value of 2.2 is derived, it is clear that the kinematics of the inner disk of IC 10 can be reproduced without the need of a dark matter halo. In this case, a M/L of 0.17 $\pm$ 0.08 is obtained. This value is closer to a M/L of 0.2 expected at the lower end for most dwarf galaxies as derived by \citet{2016AJ....152..157L}. It is important to note that this result does not exclude the possibility that dark matter may be present on larger scales in IC 10. \citet{2015AJ....149..180O} derived their mass model fit of IC 10 using high resolution VLA data. By fitting the inner disk out to $ \lesssim$ 0.57 kpc (see black arrow in Figure \ref{Fig:g}), they derived lower $\chi^{2}$ values of 0.23 and 0.07 for the NFW and ISO respectively with the stellar component dominating the gravitational potential at most radii. Comparing with the results of other BCDs from the literature, \citet{2012A&A...537A..72L} and \citet{2012A&A...544A.145L} showed that baryons may dominate the gravitational potential in the inner regions of UGC 4483 and I ZW 18, with the stellar component contributing $\sim$ 50$\%$ of the observed rotation curve of UGC 4483. This could suggest, as from our result of IC 10, that the kinematics of the inner disks of most BCDs can be described without a dark matter halo. \\

\begin{table}
\scriptsize
\captionsetup{width=7.5cm}
\caption{\small Results for the Mass Models of IC 10. The densities have the units of 10$^{-3}$M$_{\odot}$pc$^{-3}$ and the radii are in units of kpc.}
\begin{minipage}{\textwidth}
\begin{tabular}{l@{\hspace{0.4cm}}c@{\hspace{0.4cm}}c@{\hspace{0.4cm}}c@{\hspace{0.4cm}}c@{\hspace{0.4cm}}}   
\hline

%Model &Parameter & Result &sa &fr\\
                         %&(sec)       &(Mpc) & (mag) & ($L_{\odot}$) &($\rm M_{\odot}yr^{-1}$)  \\
                %~~~~~~(1)    &   (2)        &  (3)   \\         
   
\hline \hline  
%\multicolumn{6}{@{} p{8.5 cm} @{}}{\footnotesize{\hspace{3cm} VLT/NACO DATA}}\\\\
%&&IC 10&& \\
%\cline{2-3}
Model&&values& \\ 
\cline{2-4}
%ISO &&&&\\ 
\cline{2-3}\\
ISO&$\Upsilon_{*}$ & r$_{c}$ &$\rho_{0}$ &$\chi_{\text{red}}^{2}$\\ \\
& 0.04 $\pm$ 0.04 &0.21 $\pm$ 0.03&476.33 $\pm$ 230.02&1.00\\

%&&NFW Halo&& \\
%\cline{2-3}
%NFW&&&&\\
\\ \hline
NFW &$\Upsilon_{*}$ & c &R$_{200}$ &$\chi_{\text{red}}^{2}$\\ \\
& 0.11 $\pm $ 0.02 & 0.03$\pm$ 186.95 & 852.42 $\pm$ 205987.91& 1.60\\ \hline

no-DM&$\Upsilon_{*}$&&&$\chi_{\text{red}}^{2}$\\ \\
&0.17 $\pm$ 0.08&&&2.2\\ \hline

%\multicolumn{6}{@{hljvghcvgyc} p{8.5 cm} @{}}{umhgvu}\\\\

\end{tabular}   
\end{minipage}
\label{tab:template4}
\end{table}  

\section{Numerical Simulations} \label{s9}

The origin of the stream-like features seen in IC 10 has been 
widely discussed (see \citet{2013ApJ...779L..15N, 2014AJ....148..130A} and 
references therein). In general, these discussions focus on whether they are caused by an interaction with M~31
or whether they arise from dwarf-dwarf interactions or some other trigger. In particular \citet{2013ApJ...779L..15N}
compared the Northern Extension to 1000 possible 
orbits and concluded that the origin of that 
feature is not due to an interaction with M~31.
However, streams do not
quite follow the orbits of their 
progenitors \citep{Sanders2013}.  Thus, it is 
possible that the separation of the Northern Extension
from the possible orbits of IC 10 in Fig. 2 of 
\citet{2013ApJ...779L..15N} may be due to the 
tidal evolution.  To test this possibility for 
both the Southern Plume and the Northern Extension,
we have run a suite of simulations of possible 
IC~10 -- M~31 interactions.

The parameter space of the IC~10 -- M~31 system is quite large. It includes the individual galaxy 
properties as well as the relative geometry between the two systems. While the system has
been studied in detail, the uncertainties in these parameters allow for quite different 
orbital histories. In this work we have focused on the effects of uncertainties in IC ~10's proper motions.
For simplicity, we run nine simulations, where the only difference is the proper motion vector.  While these
are representative of the orbits allowed by IC~10's tangential motion, it is important to 
be cautious when interpreting the results of these simulations.

We have generated the initial conditions (ICs) for each galaxy using the 
\textsc{GalactICS} code \citep{Kuijken1995,Widrow2008,Deg2018}. The
version of the code given in \citet{Deg2018} can generate 
equilibrium models with a bulge, two stellar disks, 
a gas disk, and a dark halo. Once the ICs for M~31 and IC~10 are determined, 
the combined system is evolved using the \textsc{Gadget-2} smoothed
particle hydrodynamics code \citep{Springel2005}.

\subsection{Models}

The \textsc{GalactICS} code generates equilibrium models using a S\'{e}rsic bulge,
two exponential stellar disks, an exponential gaseous disk, and a 
double power law halo.  \citet{Prugniel1997} found 
that the density that gives rise to a S\'{e}rsic 
profile is
\begin{equation}
\rho_{b}(r)= \rho_{0,b}\left(\frac{r}{r_{b}}\right) e^{\left[-b\left(\frac{r}{r_{b}}\right)^{1/n}\right]},
\end{equation}
where $r$ is the spherical radius and $p = 1 - 0.6097/n +
0.05563/n^{2}$.  However, \textsc{GalactICS} parameterizes the bulge
with a scale velocity, $\sigma_{b}$, given 
by
\begin{equation}
\rm \sigma_{b}= [4\pi n b^{n(p-2)}(n(2-p))r^{2}_{b}\rho_{b}]^{1/2}.
\end{equation}
The exponential stellar disks have a density of 
 \begin{equation}
\rm \rho_{d}(R,z)= \frac{M_{d}}{4 \pi R_{d}^{2} z_{d}} 
e^{(-R/R_{d})}\textrm{sech}^{2}(z/z_{d}) C(R;R_{t},\delta R_{t}),
\end{equation}
in cylindrical coordinates where $M_{d}$ is the disc 
mass, $R_{d}$ is the disc scale length, z$_{d}$ is the disc scale height,
and $C(R;R_{t},\delta R_{t})$ is a truncation factor given by
\begin{equation}
C(R;R_{t},\delta R_{t})=\frac{1}{2}\textrm{erfc}\left(\frac{R-R_{t}}{\sqrt{2}\delta R_{h}} \right).
\end{equation}
The double-power law halo density is
\begin{equation}
\rm \rho(R) = \frac{2^{2-\alpha}\sigma_{h}^{2}}{4 \pi r_{h}^{2}}
\frac{1}{u^{\alpha}(1+u)^{\beta-\alpha}}C(r;r_{t},\delta r_{t})~,
\end{equation}
where $\sigma_{h}$ is a scale velocity, $u=r/r_{h}$, $r_{h}$ is a 
scale radius, $\alpha$ and $\beta$ are the inner and outer slopes 
respectively, $C(r,r_{t},\delta r_{t})$ is another truncation factor.

M 31 is the most well studied galaxy after the MW. As such, it has been constrained quite well 
by a variety of observations. We have chosen to use the isothermal model found 
by \citet{Chemin2009} to build the M 31 ICs. The specific parameters of this model
are listed in Table \ref{tab:GalTab}. For IC 10, we 
modified the NFW mass model to a \textsc{GalactICS}
analogue with the same general rotation curve,
gas mass, and scale lengths.
The parameters for this model are 
also listed in Table \ref{tab:GalTab}.

\begin{table*}
 \small
 \caption{Summary of the model parameters.}
 \label{tab:GalTab}
 \begin{center}
 \begin{tabular}{llll}
  \hline
 Parameter & Units & M 31& IC 10  \\
 \hline
Halo scale velocity: $\sigma_{h}$ & km s$^{-1}$ &390 & 31\\
Halo scale radius: $R_{h}$& kpc& 5.1& 2\\
 Inner slope:$\alpha$& -&0 &1 	\\
 Outer slope:$\beta$ &- & 3&3\\
 Halo Truncation: $r_{t}$ & kpc & 100& 8\\
 Halo Truncation Length: $\delta r_{t}$ & kpc &20 & 3 \\
 Disk mass:$M_{d}$ & 10$^{9} \ M_{\odot}$ & 59. & 0.12\\
 Disk scale length: $R_{d}$ & kpc& 5.6& 0.3\\
 Disk truncation: $R_{t}$ & kpc &25 & 1.1\\
 Disk truncation length: $\delta R_{t}$ & kpc & 3& 0.1\\
 Gas mass:$M_{g}$ & 10$^{9} \ M_{\odot}$ & 10. & 0.12\\
 Gas scale length: $R_{g}$ & kpc &8.5 & 0.9\\
 Gas Temperature: $T$ & K & $10^{4}$ & $10^{4}$\\
 Gas trunctation: $R_{g,t}$ & kpc & 40& 4\\
  Gas trunctation length: $\delta R_{g,t}$ & kpc & 2&0.9 \\
Bulge S\'ersic index: n& - & 1.1 &- \\
  Characteristic bulge velocity:$\sigma_{b}$& km s$^{-1}$ & 250& -\\
Bulge scale length:$R_{b}$ & kpc & 1.3 & -\\
  \hline
 \end{tabular}
   \end{center}
\end{table*}

\subsection{Geometry}

M 31 is located at $(783 \textrm{~kpc}, 121.2^\circ, -21.6^\circ)$ with a radial velocity
of $v_{r}=-300~\textrm{km s}^{-1}$
\citep{deVaucouleurs1991}.  \citet{VanDerMarel2012} utilized HST observations to
determine that the proper motions of M 31 are $(-76.3,~ 45.1)~\textrm{km s}^{-1}$ in RA and 
Dec.  IC 10 is located at $(794 \textrm{~kpc}, ~119^\circ, ~-3.3^\circ)$ \citep{2012AJ....144..134H} with a radial
velocity of $v_{r}=-348 \textrm{km s}^{-1}$
\citep{2014AJ....148..130A}.  Unlike many dwarf galaxies, there are 
measurements of IC 10's proper motions.  \citet{Brunthaler2007} used the 
Very Large Baseline Array to see the movement of the galaxy relative to 
background masers.  They found IC 10's proper motions to be
$(-122\pm31, ~97 \pm 27)~\textrm{km s}^{-1}$ in RA and Dec.

\subsection{Simulations}

Determining the initial relative phase-space coordinates for simulation of an IC~10 -- M~31 system 
that will end up with IC~10 at the current position and velocity relative to M~31 is not a trivial
task. Our solution is to essentially run each simulation twice. First, M 31 is placed at the 
origin and the model IC 10 system is placed at the current phase-space coordinates. With IC 10 orbiting around M 31, the 
system is then evolved backwards for $6$ Gyr. The center-of-mass and velocity of IC 10 in
the final snapshot of this first simulation is used as the initial phase-space coordinates
for the forwards simulation. The IC 10 system in the final snapshot of the forwards simulation
is used for our analysis. This results in a system 
that is close to the observed phase-space coordinates 
of IC 10, with some small differences caused 
by dynamical friction and the disruption of the 
IC 10 model.

Given the relatively large uncertainty in the proper motions of IC 10, we ran a set of nine simulations
exploring the range of possible tangential velocity vectors. Each simulation has
$1.6 \times10^{6}$ M 31 particles broken up into $2\times10^{5}$ gas particles, $3\times 10^{5}$ disk particles, $10^{5}$ bulge particles, and 
$10^{6}$ halo particles, and $3\times10^{5} $ IC 10 particles, with $5\times10^{4}$ gas particles,
$5\times10^{4}$ disk particles, and $2\times10^{5}$ halo particles.  Each simulation
was evolved for 6 Gyr using the \textsc{Gadget 2} code with a softening length of 0.05 kpc.  The 
evolution time was chosen to give IC 10 enough time to complete at least one orbit.

\begin{figure*}
\centering
    \includegraphics[width=90mm]{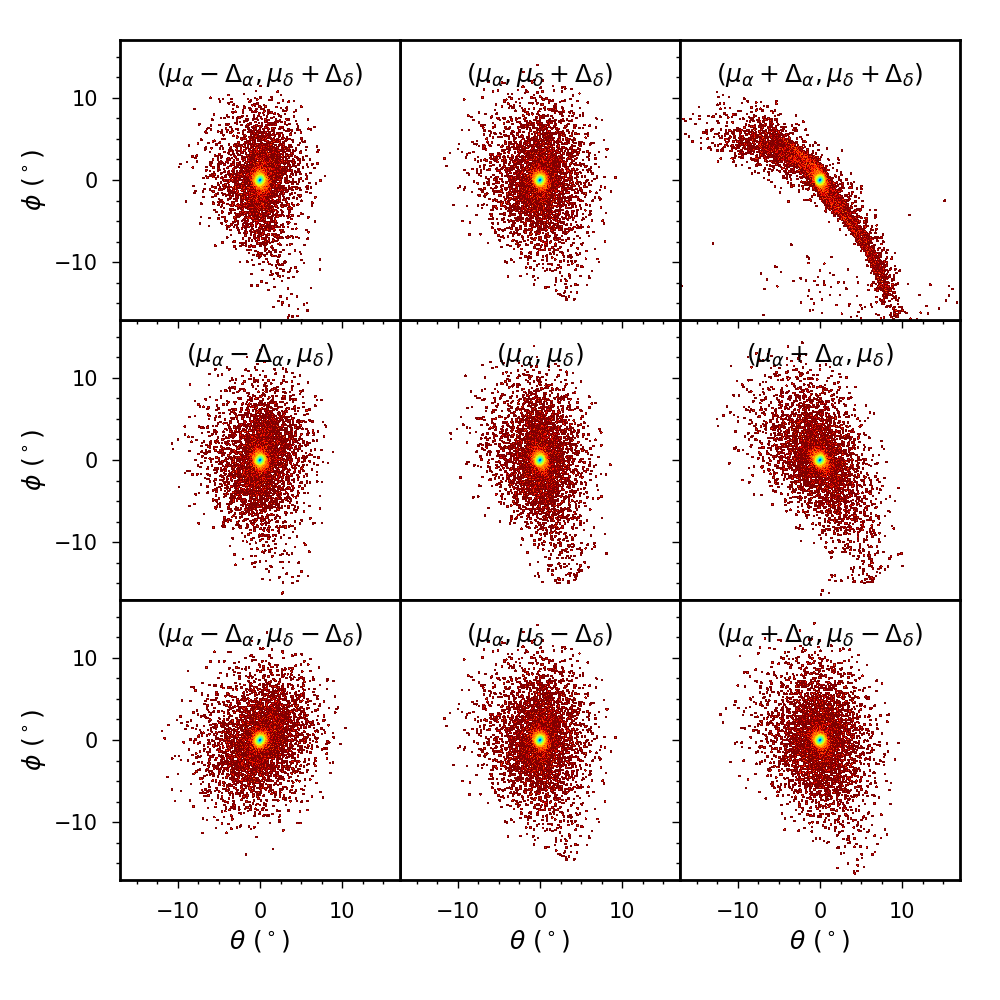} 
\caption{Mock images of the final state of the IC 10 gas disk in the nine simulations.  The labels indicate
the proper motions used to determine the initial conditions.  The maps use a logarithmic surface density.
The axes are in units of degrees.}
  \label{Fig:IC10SimSnaps}
\end{figure*}

Figure \ref{Fig:IC10SimSnaps} shows mock observations of the IC 10 gas disk at the final time step of each simulation.
These observations are made using 120 arcsec pixels with a 300 arcsec beam.  
Most of the systems do not show clear tidal features other than a cloud of disrupted
particles around the central regions. However, the simulation with 
$(\mu_{\alpha}+\Delta_\alpha,\mu_{\delta}+\Delta_\delta)$ shows a clear tidal stream.
Therefore, there are at least some possible orbits for IC 10 that can produce tidal features.  

\begin{figure}
\centering
    \includegraphics[width=70mm]{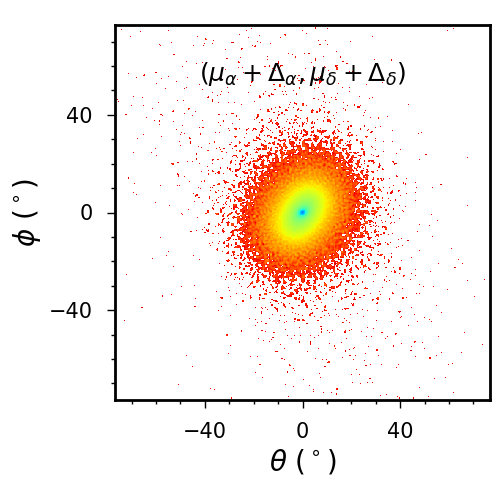} 
\caption{A mock observation of the
($\mu_{\alpha}+\Delta_\alpha,\mu_\delta+\Delta_\delta$) snapshot
using 18 arcmin pixels and truncated the 
surface density below $6.9\times10^{18}~\textrm{cm}^{-2}$.  
The axes are in units of degrees.}
  \label{Fig:IC10_Sim_+R+DZIn}
\end{figure}

The tidal feature seen in $(\mu_{\alpha}+\Delta_\alpha,\mu_{\delta}+\Delta_\delta)$ (top right of Fig. 4.12) is
quite interesting. It is much larger than the stream-like feature in the H\textsc{i} disk of 
IC 10. Moreover, it is a very low surface-density feature.  To investigate this 
in greater detail, Figure \ref{Fig:IC10_Sim_+R+DZIn} is a mock image with the same pixel scale and column 
density as the DRAO observations.  To be clear,
the image is limited to 
a column density of $6.9\times10^{18}~\textrm{cm}^{-2}$,
corresponding to a surface density of $ 0.055 ~\textrm{M}_{\odot} \textrm{pc}^{-2}$. Any 
pixel values lower than this limit are artificially set to zero. With these limits and at this scale, the image does not
show strong evidence of a tidal feature as the extended 
tidal tails lies below the detection limit.

These results suggest that it is possible that
IC 10 is on an orbit that allows for a disruption by M 31. However, the features from 
such a disruption are more extended than those observed.  Moreover they are at a lower surface density
than can be reached by current observations. Thus, it is unlikely that the actual
features seen in the H\textsc{i} disk of IC 10 are caused by the interaction with M 31 as suggested by \citet{2013ApJ...779L..15N}.

It is important to note that these conclusions have been reached using a limited
number of simulations. While these are representative of possible orbital histories 
for IC 10, there is a large area of parameter space that we have yet to explore.
We have not included the potential of M33, which may affect IC 10's orbital history.
Nor have we varied the M 31 or IC 10 models, or explored the effects of 
uncertainties in distance or radial velocities. 

\begin{figure}
\centering
    \includegraphics[width=80mm]{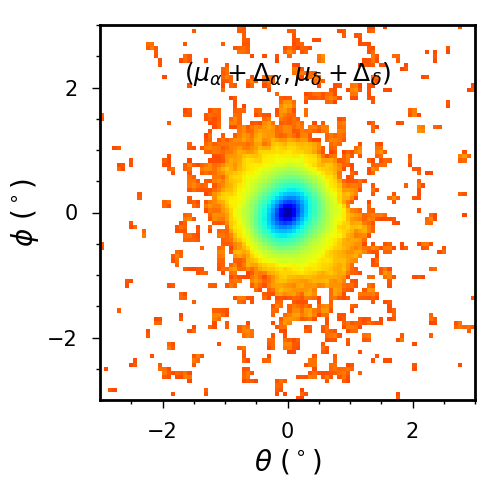} 
\caption{A mock observation of the
($\mu_{\alpha}+\Delta_\alpha,\mu_\delta+\Delta_\delta$) snapshot
using 4 arcmin pixels and truncated the 
surface density below $7\times10^{17}~\textrm{cm}^{-2}$.  
The axes are in units of degrees.}
  \label{Fig:brenda}
\end{figure}

Nonetheless, the results are quite interesting and suggest that
further observations that reach to lower column densities
be carried out. While the currently observed features are 
unlikely to be caused by M 31, the presence or absence of 
extended features will further constrain the orbital history of
IC 10. Returning to the observed features, it appears that 
a mechanism other than the IC~10 -- M~31 interaction is needed to 
explain the observed disruption in the IC~10 disk. The simulations do rule out M~31 as a source for the Southern stream. While there may be some issues with conversions, it is highly unlikely that the low column density stream of the simulation could be enhanced to such a large extent as to generate that particular feature.  However, it is possible that the stream in the simulation could be related to the Northern extension. To that end, we have reproduced the GBT limited observation, see Figure \ref{Fig:brenda}. For this new image, the pixel size is 4$^{\prime}$, the beam FWHM is 9$^{\prime}$, and the image is limited to a +/- 3 degrees about the IC 10 center. This figure does not quite
rule out a tidal origin for the Northern extension,
as the column density of the underlying stream is 
near the GBT limit.  Some other orbits might 
form streams above the limit and appear as
an extension.  In order to fully rule out an M~31 interaction
as the origin,
it will be necessary to follow 
\citet{2013ApJ...779L..15N}, and compare P-V
diagrams of a set of simulated streams to
IC~10.  It is important to use the 
stream simulations rather than the impressive
suite of orbits used in \citet{2013ApJ...779L..15N}
due to the difference between streams and orbits.

\section{Summary} \label{s66}
We have analyzed $\sim$1000 hours of DRAO H\textsc{i} observations of the blue compact dwarf galaxy IC 10 combined with simulations to study the H\textsc{i} morphology and kinematics, and investigate if an interaction of IC 10 with M 31 can reproduce the observed H\textsc{i} features we see in the outer regions of IC 10. The main outcome of this study are:
\begin{enumerate}
\item We are able to verify the faint H\textsc{i} feature previously reported from GBT observations at a  column density limit $\sim$ 8 $\times$ 10$^{17}$ cm$^{-2}$. In order to further carry out detailed kinematical analysis on the H\textsc{i} feature, there is a need for high sensitivity high resolution H\textsc{i} observations.\\

\item The compact configuration of the DRAO has allowed us to detect the extended H\textsc{i} disk of IC 10 further than previous H\textsc{i} interferometric studies. The H\textsc{i} mass of 7.8 $\times$ 10$^{7}$M$_{\odot}$ derived from our DRAO data is comparable to the H\textsc{i} mass derived from the single dish GBT data. We detect $\sim$ 33$\%$ more H\textsc{i} mass than the VLA LITTLE THINGS data.\\

\item The rotation curve of IC 10 is similar in form to those of other dwarf galaxies \citep{10.1093/mnras/stx2256}, having a linearly rising (solid body) inner portion and then turning over to become approximately flat. On the other hand, the slope of the inner part of IC 10 is $\sim$ 3 times steeper (65 km\ s$^{-1}$/kpc) than what has been derived for most dwarf galaxies. \citet{10.1093/mnras/stx2256,10.1093/mnras/sty1056} calculates the inner slopes of 21, 21, and 13 km\ s$^{-1}$/kpc for NGC 6822, Sextans A and B while \citet{1996AJ....111.1551M} and \citet{1998MNRAS.300..705M} derive the inner slopes of 25 km\ s$^{-1}$/kpc and 22 km\ s$^{-1}$/kpc for BCD galaxies NGC 2915 and NGC 1705. We derive a mean V$_{sys}$ = -351 $\pm$ 1.8 km s$^{-1}$, PA = 65 $\pm$ 4$^{\circ}$, and $i$ = 47 $\pm$ 6$^{\circ}$, these values are consistent with the literature.\\

\item We ran mass models with a dark matter halo component. A M/L of 0.04 derived from the ISO fit is too small and does not make physical sense. On the other hand, the large errors derived for the NFW fitted parameters and the nonphysical small value of the concentration parameter c suggest that this model is not suitable to explain the mass distribution of the inner disk of IC 10. Although a no dark matter model has a slightly higher reduced $\chi^{2}$, we find that we can represent the kinematics of the inner disk of IC 10 without the need of a dark matter halo, which does not exclude that dark matter may exist on a larger scale in this galaxy. \\

 %\item We ran mass models with a dark matter halo component. A M/L of 0.04 derived from the ISO fit does not make physical. On the other hand, the large errors derived for the NFW fitting parameters suggest that this model may not be suitable to explain the kinematics of IC 10. Although a no dark matter model has a slightly higher reduced $\chi^{2}$, we find that we can represent the kinematics of the inner disk without the need of a dark matter halo, which does not exclude the possible presence of dark matter on a larger scale. 

\item It is unlikely that the H\textsc{i} features observed in IC 10 are caused by an interaction with M 31. The disruptions seen in our simulations are at lower densities and larger than those observed. We think that, as suggested by \citet{1998AJ....116.2363W}, the H\textsc{i} extensions with different kinematics seen south, east and west of the main core of IC 10 may be the result of accretion, as suggested by their counter-rotation and by the increase in velocity dispersion observed at the point of contact of those regions with the inner regular disk. The accreted material might be coming from the low column density H\textsc{i} gas that is not visible to the sensitivity of our existing telescopes.\\

\end{enumerate}

\section{ACKNOWLEDGEMENT}
CC's work is based upon research supported by the South African Research Chairs Initiative (SARChI) of the Department of Science and Technology (DST), the South African Radio Astronomy Observatory (SARAO) and the National Research Foundation (NRF). The research of BN \& ND have been supported by SARChI bursaries. This work was made possible by a Brandon University Research Committee (BURC) grant to T.F. The Dominion Radio Astrophysical Observatory is a National Facility operated by the National Research Council Canada. The authors would like to sincerely thank Dr. Andrew Gray (Operations Manager of DRAO) and Dr. Roland Kothes for their unstinting efforts on our behalf in scheduling new observations for this paper, and their incredible support of graduate student projects. We would also like to thank the reviewer for their positive criticisms which motivated us to obtain more observations.
%%%%%%%%%%%%%%%%%%%%%%%%%%%%%%%%%%%%%%%%%%%%%%%%%%



%\bibliographystyle{mnras}


%%%%%%%%%%%%%%%%%%%%%%%%%%%%%%%%%%%%%%%%%%%%%%%%%%

%%%%%%%%%%%%%%%%% APPENDICES %%%%%%%%%%%%%%%%%%%%%

%\appendix



%%%%%%%%%%%%%%%%%%%%%%%%%%%%%%%%%%%%%%%%%%%%%%%%%%

% Don't change these lines

%%%%%%%%%%%%%%%%%%%%%%%%%%%%%%%%%%%%%%%%%%%%%%%%%%



%\bibliographystyle{mnras}


\end{document}